\documentclass{article}
\usepackage{subcaption}

\usepackage[main, final]{neurips_2026}

\usepackage[utf8]{inputenc} % allow utf-8 input
\usepackage[T1]{fontenc}    % use 8-bit T1 fonts
\usepackage{hyperref}       % hyperlinks
\usepackage{url}            % simple URL typesetting
\usepackage{booktabs}       % professional-quality tables
\usepackage{amsfonts}       % blackboard math symbols
\usepackage{nicefrac}       % compact symbols for 1/2, etc.
\usepackage{microtype}      % microtypography
\usepackage{xcolor}         % colors
\usepackage{graphicx}
\usepackage{algorithm}
\usepackage{algpseudocode}
\usepackage{amsmath}
\usepackage{amsthm}
\usepackage{amssymb}
\usepackage{makeidx}
\usepackage{enumitem}
\usepackage{hyperref}
\hypersetup{hidelinks}

\usepackage{titletoc} % appendix-only partial contents
\makeindex

\newtheorem{assumption}{Assumption}
\newtheorem{theorem}{Theorem}
\newtheorem{lemma}{Lemma}
\newtheorem{corollary}{Corollary}
\newtheorem{proposition}{Proposition}

\title{Epistemic Social Learning: Latent Behavioral Structure under Endogenous Multi-Agent Interaction}

\author{%
  Jainendra Shukla \qquad Dhruv Jaiswal \qquad Divyanshi Beniwal \qquad Kiriti Kanjilal \\
  \normalfont IIIT-Delhi, New Delhi, India \\
  {\small\texttt{\{jainendra,dhruv23200,divyanshi23209,kanjilal\}@iiitd.ac.in}}
}

\begin{document}

\maketitle

\begin{abstract}
In multi-agent interaction, a learner's actions shape the observations from which it infers other agents' behavior. A behavioral model may predict these observations well yet support poor adaptation when opponents change. We introduce \emph{Epistemic Social Learning} (ESL), which couples fast, interaction-specific Bayesian beliefs with slower online learning of population-shared behavioral prototypes. Belief-conditioned decisions influence subsequent observations and, in turn, the representations learned from them. Under explicit regularity and averaging assumptions, we provide a conditional stochastic-approximation characterization of these coupled dynamics. Across twelve synthetic $3\times3$ cooperative, coordination, and adversarial games, ESL achieves the lowest aggregate post-switch regret among the evaluated online latent-adaptation methods, including a switch-aware Bayesian filter. Additional experiments examine observation loss, noise, and asymmetric incentives. These findings illustrate how predictive fit can differ from decision-relevant adaptation under policy-dependent interaction.
\end{abstract}

\section{Introduction}
\label{sec:intro}

Adaptive agents interact with partners whose behavior varies and changes. Beyond predicting the next action, they must learn behavioral structure that remains useful as interactions evolve. Multi-agent observations are endogenous: beliefs shape actions, which determine observed responses and update subsequent beliefs. Thus, the learner helps generate its own training data.

This feedback creates a subtle failure mode. A behavioral representation can fit the observations generated under the current interaction policy yet omit distinctions that become important after behavior changes. For example, two latent behaviors may appear effectively identical under the actions currently taken by the learner, while implying different best responses once the interaction shifts. Predictive fit and decision-relevant adaptation can therefore come apart. This motivates our central question: \emph{how should agents learn shared latent behavioral structure when their own policies influence the evidence available for learning it?}

We propose \emph{Epistemic Social Learning} (ESL), a two-timescale framework for this setting. ESL maintains learnable, population-shared behavioral prototypes: parameterized components of a behavioral model, not prespecified ground-truth opponent types. Their parameters evolve slowly across interactions, while fast, interaction-specific Bayesian beliefs express uncertainty over these components. Agents choose belief-conditioned logit responses, creating a closed loop in which beliefs shape policies, policies generate observations, and those observations update both beliefs and the shared representation. The resulting structure allows experience to be pooled across interactions while retaining opponent-specific uncertainty.

ESL combines established Bayesian modeling and online representation learning in policy-dependent interaction. We characterize its coupled dynamics conditionally using controlled-Markov stochastic approximation, and evaluate adaptation on twelve synthetic $3\times3$ games with complementary robustness and population diagnostics.

\textbf{Contributions.}
\begin{itemize}
    \item We formulate latent behavioral learning under endogenous interaction and identify the distinction between predictive fit and decision-relevant adaptation.

    \item We introduce ESL, which couples slow online learning of shared behavioral prototypes with fast interaction-specific Bayesian inference, and provide a conditional stochastic-approximation characterization of the resulting coupled dynamics.

    \item Across the twelve-game benchmark, ESL achieves the lowest aggregate post-switch regret among the evaluated online latent-adaptation baselines, including a switch-aware Bayesian filter, while complementary experiments characterize when and under what observation conditions this adaptation advantage arises.
\end{itemize}

\section{Related Work}

\textbf{Latent behavior and opponent modeling.}
Classical Bayesian opponent modeling maintains beliefs over candidate models~\cite{ALBRECHT201866}; cognitive-hierarchy approaches describe levels of strategic reasoning~\cite{10.1162/0033553041502225,stahl1995players,stahl1994experimental}. DRON and DPIQN learn policy-conditioned opponent representations~\cite{he2016opponent,hong2018deep}, whereas EM/HMM methods represent heterogeneity through latent states~\cite{dempster1977maximum,guennouni2024impact}. These approaches distinguish belief inference from learning behavioral structure, but do not by themselves specify how an online shared library changes alongside interaction-specific beliefs. Switch-aware filtering and Bayesian online changepoint detection~\cite{adams2007changepoint} also address nonstationarity. Our SA-FP comparator uses fixed emissions; learned-emission HMMs and changepoint comparators remain future work. Related decision-dependent distribution shifts are studied in performative prediction~\cite{pmlr-v119-perdomo20a}; ESL focuses on shared behavioral prototypes and opponent-specific inference in repeated interaction.

\textbf{In-context and representation-based adaptation.}
Learned strategy spaces~\cite{li2025adaptively} and recurrent or theory-of-mind models~\cite{pmlr-v80-rabinowitz18a,mon-williams2025partner} support opponent adaptation. OEOM~\cite{jing2025open} constructs a pretrained opponent population; OMIS~\cite{jing2024opponent} uses separately trained components and decision-time search. These systems can draw on models trained before evaluation, whereas ESL learns its shared prototypes from the evaluated interaction stream. Both are applicable in principle, but a controlled comparison would need to match pretraining, inference, and decision-time resources. We make no superiority claim over these systems; resource-matched sequential comparisons remain open.

\textbf{Multi-agent learning and strategic adaptation.}
Fictitious play~\cite{brown1951iterative}, follow-the-regularized-leader (FTRL), and mirror descent~\cite{pmlr-v15-mcmahan11b} study policy-level learning in repeated interaction. Optimistic mirror descent has been analyzed for convergence under particular game assumptions~\cite{pmlr-v162-anagnostides22a}. LOLA~\cite{foerster2018learning} anticipates other agents' learning updates, while recursive-reasoning methods~\cite{wenprobabilistic,wen2021modelling} model higher-order beliefs. ESL instead couples online learning of shared behavioral prototypes with interaction-specific Bayesian inference and policy-dependent observations. Its stochastic-approximation characterization is conditional, not an equilibrium or no-regret guarantee.

\section{Preliminaries}

We consider a population of $N$ agents indexed by $[N]=\{1,\dots,N\}$ sharing a finite action set $\mathcal{A}$. The environment unfolds over rounds $t=0,1,2,\dots$. At each round, a set $\mathcal{E}_t \subset \{(i,j)\in[N]\times[N]:i\neq j\}$ of $L_t$ ordered interaction pairs is sampled uniformly without replacement. Each pair $(i,j)\in\mathcal{E}_t$ selects actions $a_i^t,a_j^t\in\mathcal{A}$ and receives payoffs from a stage game $u:\mathcal{A}\times\mathcal{A}\to\mathbb{R}$.

Within each round, encounters are processed sequentially using the live, in-place belief and prototype states. A selected ordered pair $(i,j)$ designates $i$ as the observing learner; the reverse belief of $j$ updates only on a separately selected directed observation.

Agents reason about opponents through a shared latent behavioral space updated jointly with interaction dynamics. The interaction set $\mathcal E_t$ determines \emph{which ordered pairs interact}, whereas for each selected $(i,j)\in\mathcal E_t$ the observation indicator $W_{ij,t}\in\{0,1\}$ separately specifies \emph{whether agent $i$ receives an informative signal} about $j$. In particular, an interaction can occur without a subsequent belief update when $W_{ij,t}=0$. Let $\mathcal Y$ be the finite signal space and $P:\mathcal A\to\Delta(\mathcal Y)$ the observation channel, where $P(s\mid a)$ is the probability of observing $s\in\mathcal Y$ given the opponent action $a\in\mathcal A$; the conditional signal law is $s_{ij,t}\sim P(\cdot\mid a_j^t)$ when $W_{ij,t}=1$. In the primary noiseless setting, $\mathcal Y=\mathcal A$ and $s_{ij,t}=a_j^t$. The mask may be directional: $W_{ij,t}$ need not equal $W_{ji,t}$.

We posit $K$ learnable, population-shared behavioral prototypes indexed by $k\in[K]$, each parameterized by $\theta_k\in\Theta\subset\mathbb{R}^{|\mathcal{A}|}$ and inducing an action distribution
$
\pi_k(a\mid\theta_k)=\mathrm{softmax}(\theta_k)[a].
$
Each prototype is a component of the shared behavioral model whose parameters are learned from interaction data, not a known or immutable ground-truth opponent type. The induced signal likelihood is
$
L_k(s\mid\theta_k)
=
\sum_{a\in\mathcal{A}} P(s\mid a)\pi_k(a\mid\theta_k),
$
which reduces to $\pi_k(a\mid\theta_k)$ in the noiseless setting. We denote the time-indexed collection of prototype parameters by $\Theta_t=\{\theta_{k,t}\}_{k\in[K]}$; when indexed by slow prototype updates, the same collection is written $\Theta_m$.

Each agent $i$ maintains, for every opponent $j\neq i$, a belief vector $b_{i\to j,t}\in\Delta_K^\delta$, where
$
\Delta_K^\delta
=
\left\{
b\in\mathbb{R}^K:
\sum_{k=1}^K b_k=1,\;
b_k\ge\delta
\right\},
$
for some $0<\delta<1/K$. Here $\Pi_{\Delta_K^\delta}(x):=\arg\min_{b\in\Delta_K^\delta}\|b-x\|_2^2$ is the Euclidean projection of a normalized belief vector $x$ onto the truncated probability simplex. Projection onto $\Delta_K^\delta$ ensures strictly positive beliefs and compactness. The global belief state is $B_t=\{b_{i\to j,t}\}_{i\neq j}$. Within a slow-update batch, $\Theta_m$ is held fixed, whereas $B_t$ continues to change after observed interactions.

When agent $i$ interacts with agent $j$, it forms the predictive distribution
\[
\hat{\pi}_{j\mid i,t}(a_j)
=
\sum_{k=1}^K
b_{i\to j,t}[k]\,
\pi_k(a_j\mid\theta_k),
\qquad a_j\in\mathcal{A},
\]
where $a_j$ denotes a generic candidate action (not the realized time-$t$ sample); after actions are sampled, $\hat\pi_{j\mid i,t}(a_j^t)$ denotes the probability that the pre-observation prediction assigns to the realized action $a_j^t$. Agent $i$ then samples a belief-conditioned logit best response,
\[
\pi_i(a_i\mid b_{i\to j,t},\Theta_t)
\propto
\exp\!\left(
\lambda
\sum_{a_j\in\mathcal{A}}
u(a_i,a_j)\,
\hat{\pi}_{j\mid i,t}(a_j)
\right).
\]
The inverse temperature $\lambda\ge0$ controls the randomness of this
belief-conditioned action selection: $\lambda=0$ yields a uniform
policy, and increasing $\lambda$ concentrates probability on
higher-utility actions (approaching a best response as
$\lambda\to\infty$). It does not appear directly in the Bayesian
belief or prototype update equations, but it may affect their future data
indirectly by changing the policy-dependent observations. Agent $j$ acts symmetrically, using its own belief $b_{j\to i,t}$ to form a prediction of $i$'s behavior and sample its logit response. The realized action $a_j^t$ then determines the signal $s_{ij,t}\sim P(\cdot\mid a_j^t)$ observed by $i$, which is used in the Bayesian update of $b_{i\to j,t}$. Thus both agents' belief-conditioned policies jointly determine the actions and observations that drive subsequent inference, making the signal process endogenous to the current belief state.

\section{Method: Epistemic Social Learning}
\begin{figure*}[t]
    \centering
    \includegraphics[width=\textwidth,trim=0 0 0 105bp,clip]{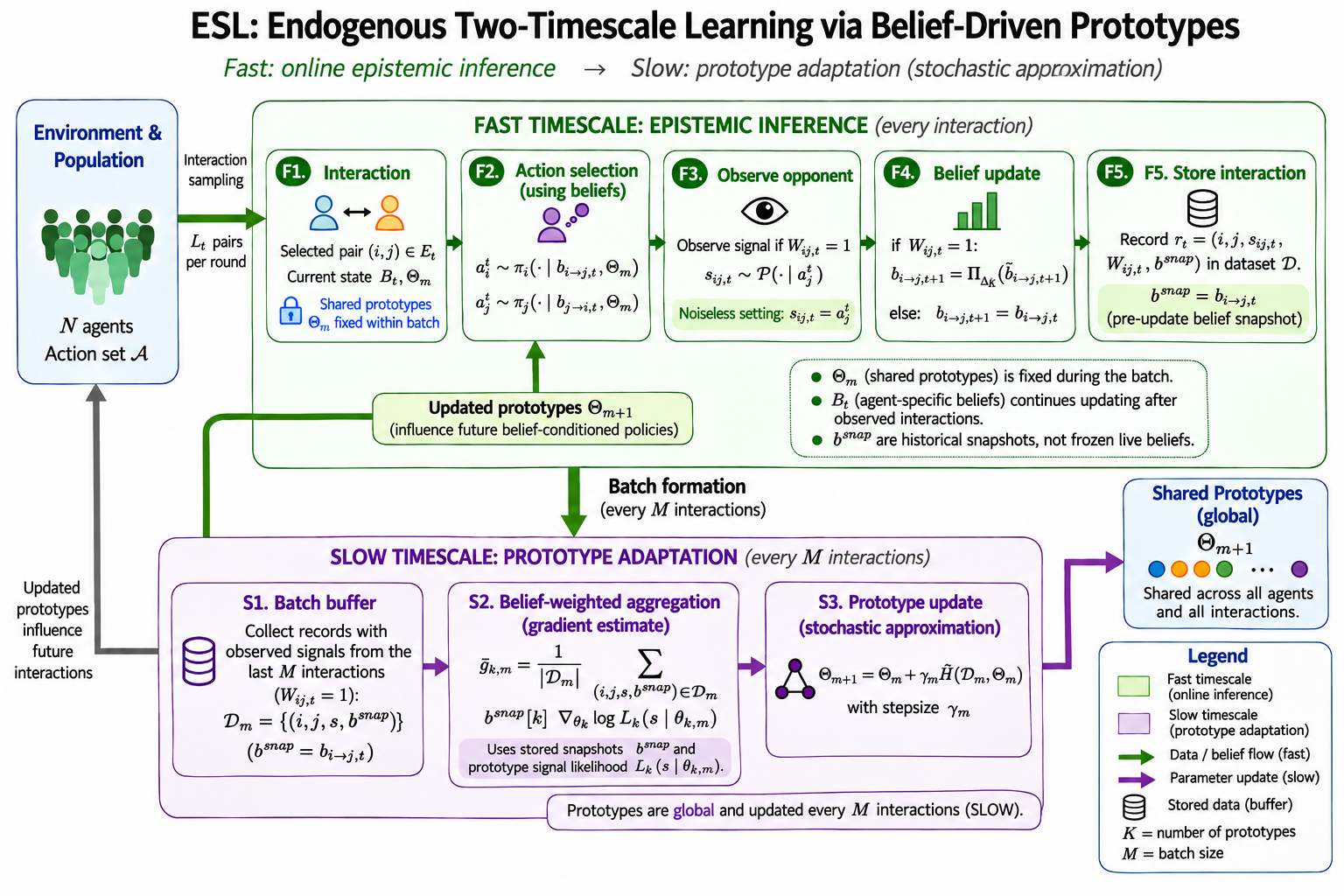}
    \caption{
    \textbf{ESL: Two-timescale belief-driven prototype learning.}
    At the fast timescale (top), agents maintain pairwise beliefs over shared prototypes and update them when opponent signals are observed. Only observed signals enter the prototype-learning buffer, together with the corresponding pre-update belief snapshots.
    At the slow timescale (bottom), batches of $M$ observed records yield belief-weighted gradients for the shared prototype update.
    This induces a feedback loop: beliefs shape interaction behavior and data generation, while accumulated evidence updates the shared prototypes, which in turn influence future beliefs and interactions.
    }
    \label{fig:esl_overview}
\end{figure*}

ESL decomposes the circularity of social learning into two coupled timescales: fast epistemic belief inference per interaction, and slow shared prototype adaptation across interactions. Intuitively, beliefs can respond immediately to each newly observed opponent signal, whereas the shared prototype dictionary aggregates belief-weighted evidence across several interactions before its next update. The two timescales are an architectural separation of update frequencies; a fixed small batch with constant step sizes is not automatically an asymptotic two-timescale convergence regime. The procedure is illustrated in Figure~\ref{fig:esl_overview} and detailed in Algorithm~\ref{algo}.

\vfill

\begin{algorithm}[H]
\caption{Epistemic Social Learning (ESL; evaluated encounter-batch loop)}
\label{algo}
\begin{algorithmic}[1]
\State \textbf{Initialize:} beliefs $B_0$, shared prototypes $\Theta_0$, buffer $\mathcal D\leftarrow\emptyset$
\For{$t=1,\dots,T$}
    \State Sample interaction set $\mathcal E_t$
    \For{each $(i,j)\in\mathcal E_t$}
        \State Sample $a_i^t\sim\pi_i(\cdot\mid b_{i\to j,t},\Theta_t)$ and $a_j^t\sim\pi_j(\cdot\mid b_{j\to i,t},\Theta_t)$
        \State $b^{\mathrm{snap}}\leftarrow b_{i\to j,t}$; sample observation mask $W_{ij,t}$
        \If{$W_{ij,t}=1$}
            \State Observe $s_{ij,t}$; update $b_{i\to j,t+1}$ via \eqref{eq:belief-update}
            \State Append $(i,j,s_{ij,t},b^{\mathrm{snap}})$ to $\mathcal D$
        \Else
            \State $b_{i\to j,t+1}\leftarrow b^{\mathrm{snap}}$
        \EndIf
        \If{$|\mathcal D|=M$}
            \State Update shared $\Theta$ via \eqref{eq:prototype-update} using $\mathcal D$; clear $\mathcal D$
        \EndIf
    \EndFor
\EndFor
\end{algorithmic}
\end{algorithm}

\subsection{Fast Timescale: Epistemic Inference under Endogenous Interaction}

%In interactive settings, the data distribution is shaped by the agents’ own policies, which evolve as beliefs are updated over time. Consequently, models fitted to historical observations may become misaligned with the current interaction dynamics. Moreover, since each agent’s behavior depends on its beliefs about others, interactions induce a feedback loop in which agents continuously adapt to one another, a phenomenon closely related to belief-based and recursive reasoning in multi-agent systems \cite{camerer2004cognitive, albrecht2021multi}.

%To address this, ESL performs online belief-based inference at the level of individual interactions. Beliefs are updated using signals generated by the interaction process itself, yielding a streaming form of Bayesian filtering over latent behavioral types \cite{doucet2001sequential}. Because the update is followed by projection onto \(\Delta_K^\delta\), it is a regularized variant of Bayesian filtering under interaction-driven data operating under a non-stationary, policy-dependent data distribution induced by the interaction dynamics.

For each observed interaction $(i,j)\in\mathcal{E}_t$ with $W_{ij,t}=1$, agent $i$ performs a projected Bayesian update:
\begin{equation}
\begin{aligned}
\tilde b_{i\to j,t+1}[k]
&\propto
b_{i\to j,t}[k]\,
L_k(s_{ij,t} \mid \theta_{k,t}), \\
b_{i\to j,t+1}
&=
\Pi_{\Delta_K^\delta}\big(\tilde b_{i\to j,t+1}\big)
\end{aligned}
\label{eq:belief-update}
\end{equation}
where the proportionality denotes normalization over $k$. Beliefs corresponding to unobserved interactions remain unchanged. Projection onto $\Delta_K^\delta$ prevents boundary degeneracy and confines beliefs to a compact interior simplex. Boundedness of the prototype likelihood scores is a separate condition, stated in Assumption~\ref{assum:smoothness}.

\subsection{Slow Timescale: Prototype Adaptation under Endogenous Feedback}

%The belief-based inference process described above yields a collection of interaction-specific opponent models, each induced by the belief state at the time of observation. Since these models are formed under evolving interaction dynamics, any single belief-induced model may provide a partial or noisy explanation of behavior. To obtain a robust and transferable representation, ESL aggregates these models into a shared set of prototype parameters.

%This aggregation operates at a slower timescale and integrates information across interactions through belief-weighted updates. Let $\mathcal{D}_m$ denote the set of interaction records collected over a batch of rounds, where each record consists of $(i,j,s,b^{\mathrm{snap}})$ and $b^{\mathrm{snap}}$ is the belief at the time of observation. The prototype parameters are updated as

At the slower timescale, ESL aggregates interaction records 
$\mathcal{D}_m = \{(i,j,s,b^{\mathrm{snap}})\}$ collected under 
frozen $\Theta_m$ into a shared prototype update. Here $b^{\mathrm{snap}}$
is agent $i$'s belief about $j$ immediately before incorporating signal $s$;
unobserved interactions do not contribute a signal-likelihood score:

\begin{equation}
\begin{aligned}
\bar g_{k,m}
&=
\frac{1}{|\mathcal{D}_m|}
\sum_{(i,j,s,b^{\mathrm{snap}})\in \mathcal{D}_m}
b^{\mathrm{snap}}[k]\,
\nabla_{\theta_k} \log L_k(s \mid \theta_k), \\
\theta_k
&\leftarrow
\theta_k + \gamma_m
\left(
\bar g_{k,m} - \eta_{\mathrm{reg}} \theta_k
\right)
\end{aligned}
\label{eq:prototype-update}
\end{equation}

%where $\eta_{\mathrm{reg}} > 0$ is an $\ell_2$ regularization coefficient ensuring stability of the prototype parameters. Let $m$ index the slow-timescale prototype updates, corresponding to interaction steps $t \in [mM, (m+1)M - 1]$, during which $\Theta_m$ remains fixed. The slow-timescale update bears structural resemblance to online EM \cite{cappe2009line}. However, unlike EM, the weighting distribution is not a posterior over a fixed data distribution; it is induced by a policy-dependent interaction process. This distinction fundamentally changes the objective being optimized. Because both the weights $b^{\text{snap}}$ and the observed signals $s$ depend on the evolving global belief state $B_t$, the noise in the gradient estimate is state-dependent, defining a controlled Markovian noise process rather than i.i.d.\ sampling. As a result, the effective data distribution is policy-dependent and non-stationary, violating the i.i.d. assumptions underlying EM-style approaches.

\paragraph{Implication of endogenous coupling.}
Because interaction policies can change the data distribution, standard exogenous or i.i.d.\ sampling assumptions need not hold. In particular, predictive likelihood on the realized interaction stream need not align with decision quality under behavioral change. The following proposition formalizes this mismatch.

\begin{proposition}[Predictive fit need not imply post-shift adaptation]
\label{prop:likelihood-mismatch}
There exists a finite two-action, two-type interaction process with policy-dependent information acquisition and two admissible behavioral representations $\Theta^{\mathrm L}$ and $\Theta^{\mathrm A}$ such that, for some $\xi>0$ and sufficiently large pre-switch horizon $T$,
\[
\mathbb P\!\left(
\mathcal L_T^{\mathrm{cl}}(\Theta^{\mathrm A})
<
\mathcal L_T^{\mathrm{cl}}(\Theta^{\mathrm L})
\right)
\ge
1-2e^{-2\xi^2T},
\]
while, for a sufficiently long post-switch evaluation window,
\[
\mathbb E[\mathrm{PSR}(\Theta^{\mathrm L})]
>
\mathbb E[\mathrm{PSR}(\Theta^{\mathrm A})].
\]
Here $\mathcal L_T^{\mathrm{cl}}$ is the closed-loop predictive log-score defined in Appendix~\ref{app:likelihood-adaptation-mismatch}. Thus, better predictive fit on the realized interaction stream need not imply better subsequent adaptation.
\end{proposition}

The construction is existential: it does not claim that endogeneity is necessary for the mismatch or that ESL necessarily learns the separated representation. Appendix~\ref{app:likelihood-adaptation-mismatch} gives the construction, sufficient separation conditions, and a matched belief-independent control.

%Proposition~\ref{prop:likelihood-mismatch} formalizes a key failure mode of likelihood-based learning under endogenous interaction: optimizing predictive fit on the realized data stream can degrade the ability to adapt under behavioral shifts. 

%ESL mitigates this failure by coupling belief-driven interaction with representation learning. Because beliefs influence policies, and policies determine which observations are generated, ESL preserves decision-relevant distinctions in the data-generating process rather than collapsing them under likelihood maximization. This ensures that learned prototypes remain informative for adaptation, not just prediction.

%Within each batch, the prototype parameter $\Theta_m$ remains fixed, so the fast belief process evolves under a time-homogeneous Markov kernel $P_{\Theta_m}$ over the entire batch. Consequently, mixing occurs under a stationary kernel within each batch, eliminating the need for Poisson equation analysis required in simultaneous-update stochastic approximation. The convergence analysis therefore reduces directly to the controlled Markov framework of \cite{borkar2006stochastic}.

%Because both the weights $b^{\mathrm{snap}}$ and the observed signals $s$ depend on the evolving global belief state $B_t$, the noise in the gradient estimate is state-dependent. The slow-timescale update therefore aggregates information across the invariant regimes of the fast belief dynamics rather than over an i.i.d. sample stream.

Unlike online EM~\cite{cappe2009line}, the weights $b^{\mathrm{snap}}$ in ESL arise from a policy-dependent interaction process induced by the agents' beliefs. Proposition~\ref{prop:likelihood-mismatch} shows that predictive fit on such a realized interaction stream can rank behavioral representations differently from post-shift adaptation quality. ESL couples prototype learning to belief-weighted interaction data generated under the current policies, but the proposition does not establish that ESL necessarily learns or preserves the separated representation; we assess that question empirically through prototype-recovery experiments and matched ablations. Fixed-K methods use a shared latent representation but do not adapt prototypes online.

\subsection{Two-Timescale Dynamical System Formulation}

ESL couples fast belief inference with slower shared-prototype updates. We analyze its policy-dependent stochastic dynamics through controlled-Markov approximation~\cite{borkar2008stochastic,kushner2003stochastic} and differential inclusions~\cite{benaim1999dynamics}.

For the conditional analysis, we use a homogeneous frozen-parameter Markov model for the observed-record-indexed belief process on $\mathcal B=(\Delta_K^\delta)^{N(N-1)}$. Such a kernel requires an appropriate sampling model and does not follow from fixing $\Theta$ alone under arbitrary switching schedules.

%We analyze this coupled system under a quasi-static approximation. Since prototype parameters evolve on the slower update scale $\gamma_m$, they are treated as effectively constant relative to the fast belief dynamics. For fixed $\Theta$, the fast process is therefore characterized through its invariant regimes, which induce the set-valued mean-field map $\mathcal{G}(\Theta)$ used in the next section. 

On the slower timescale, prototype parameters are updated using stochastic gradients based on interaction data, as defined in \eqref{eq:prototype-update}. Let the following denote the instantaneous update contribution.
\begin{equation}
H_k(B, i, j, s; \Theta)
=
b_{i\to j}[k]\,
\nabla_{\theta_k}\log L_k(s \mid \theta_k)
-
\eta_{\mathrm{reg}} \theta_k
\label{eq:drift}
\end{equation}
Write $H(B,i,j,s;\Theta):=(H_k(B,i,j,s;\Theta))_{k=1}^K$ for the full prototype-update vector.

The resulting recursion can be written as a stochastic approximation scheme with controlled Markov noise:
\begin{equation}
\Theta_{m+1}
=
\Theta_m + \gamma_m \, \widehat{H}(\mathcal{D}_m, \Theta_m),
\label{eq:sa-recursion}
\end{equation}
where \(\widehat{H}\) is the empirical average of the stacked vector \(H\) over observed interaction records collected while \(\Theta_m\) is fixed. In particular, the randomness in the slow update is induced by the fast belief process rather than by an exogenous i.i.d.\ data source.

%Under standard stochastic approximation arguments, the limiting behavior of the slow-timescale iterates is governed by averaged update directions induced by invariant regimes of the fast belief process. In particular, this leads to a set-valued mean-field map of the form
%\begin{equation}
%\mathcal{G}(\Theta) = \{ g_\mu(\Theta) : \mu \in \mathcal{I}(\Theta) \},
%\label{eq:mean-field}
%\end{equation}
%where \(\mathcal{I}(\Theta)\) denotes the set of invariant probability measures of the fast belief dynamics. Since the fast process has no auxiliary control variable beyond the frozen parameter \(\Theta\), the relevant ergodic occupation measures reduce here to invariant probability measures on \(\mathcal{B}\) \cite{borkar2006stochastic}. This characterization underlies the differential inclusion analyzed in the next section.

\begin{proposition}[Conditional frozen-batch Markov representation]
\label{prop:frozen-batch}
Suppose the belief process indexed by observed interaction records has a time-homogeneous Markov kernel under fixed prototype parameters. Then holding $\Theta_m$ fixed throughout batch $m$ gives a frozen kernel $P_{\Theta_m}$. This condition need not hold under an arbitrary nonstationary switching or sampling schedule; fixing $\Theta_m$ alone does not ensure it or establish mixing. Invariant-measure averaging and its finite-batch error remain subject to Assumptions~\ref{assum:fba} and~\ref{assum:regularity}.
\end{proposition}

\section{Theoretical Analysis: Convergence under Endogenous Feedback}
\label{sec:theory}

We analyze the limiting behavior of the slow-timescale prototype dynamics under endogenous, belief-coupled interaction. The analysis relies on the following assumptions.

\begin{assumption}[Smoothness and bounded scores]
\label{assum:smoothness}
For each prototype $k$, the likelihood $L_k(s\mid\theta_k)$ is
continuously differentiable in $\theta_k$, and the score function $\nabla_{\theta_k}\log L_k(s\mid\theta_k)$ is Lipschitz and uniformly bounded on the parameter region visited by the iterates.
\end{assumption}

\begin{assumption}[Step-size conditions]
\label{assum:stepsize}
The sequence $\{\gamma_m\}$ satisfies $\sum_{m=0}^\infty \gamma_m = \infty$, $\sum_{m=0}^\infty \gamma_m^2 < \infty$, and is eventually non-increasing.
\end{assumption}

\begin{assumption}[Conditional second-moment bound]
\label{assum:moment}
The martingale difference sequence $\{M_{m+1}\}$ satisfies $\mathbb{E}[\|M_{m+1}\|^2 \mid \mathcal{F}_m] \le C(1+\|\Theta_m\|^2)$ for some $C < \infty$.
\end{assumption}

\begin{assumption}[Finite-batch averaging]
\label{assum:fba}
There exists a measurable selection $\xi_m \in \mathcal{G}(\Theta_m)$ such that the tracking bias $\beta_m :=
\mathbb{E}[\widehat{H}_m \mid \mathcal{F}_m] - \xi_m$ satisfies
\[
  \lim_{m\to\infty}\;
  \sup_{n \ge m}\;
  \left\|\sum_{j=m}^{n}\gamma_j\,\beta_j\right\| = 0
  \quad\text{almost surely.}
\]
\end{assumption}

\begin{assumption}[Regularity of the mean-field map]
\label{assum:regularity}
For each $\Theta$, the fast belief process admits at least one
invariant probability measure, $\mathcal{I}(\Theta) \ne \varnothing$. The correspondence $\Theta \mapsto \mathcal{I}(\Theta)$ has non-empty compact values and is upper hemicontinuous. The induced set-valued mean-field map $\mathcal{G}(\Theta)$ is upper semicontinuous with non-empty, convex, and compact values.
\end{assumption}

The bounded-score and conditional-moment assumptions are standard for the bounded categorical-softmax setting. By contrast, Assumption~\ref{assum:stepsize} requires diminishing steps and excludes the headline constant-step implementation.

Assumption~\ref{assum:fba} is substantive: fixing prototypes within batches does not by itself establish a homogeneous observed-record kernel, joint-belief mixing, or negligible averaging bias. Appendix~\ref{app:convergence} supplies a sufficient law-level mixing condition yielding $\|\beta_m\|=O(M_m^{-1})$ and growing-batch schedules; this bound alone cannot verify the assumption at fixed $M$.

Assumption~\ref{assum:regularity} ensures the invariant-measure correspondence and averaged drift vary regularly with $\Theta$, allowing multiple invariant regimes. The resulting theorem is conditional, not an unconditional convergence guarantee.

\begin{lemma}[Boundedness of iterates]
\label{lem:boundedness}
Under Assumptions~\ref{assum:smoothness} and~\ref{assum:stepsize},
the update~\eqref{eq:prototype-update} with
$\eta_{\mathrm{reg}}>0$ and finite initialization produces an almost
surely bounded sequence $\{\Theta_m\}_{m\ge0}$.
\end{lemma}

Lemma~\ref{lem:boundedness} follows directly from the bounded score and the contraction induced by positive $\ell_2$ regularization; see Appendix~\ref{app:convergence}.

\begin{theorem}[Asymptotic characterization of prototype dynamics]
\label{thm:main-convergence}
Under Assumptions~\ref{assum:smoothness}--\ref{assum:regularity} and
Lemma~\ref{lem:boundedness}, for the homogeneous observed-record-indexed
fast-process model of Proposition~\ref{prop:frozen-batch}, the piecewise-linearly interpolated
process $\Theta(\tau)$ is almost surely an asymptotic pseudotrajectory
(APT) of the differential inclusion
\begin{equation}
  \dot{\Theta} \in \mathcal{G}(\Theta).
  \label{eq:diff-inclusion}
\end{equation}
Further, the limit set of $\{\Theta_m\}$ is almost surely a compact internally chain transitive (ICT) set
of~\eqref{eq:diff-inclusion}.
\end{theorem}

The proof places the ESL recursion within the controlled Markov stochastic approximation framework of~\cite{borkar2006stochastic}; the APT and ICT conclusions then follow from standard differential inclusion arguments~\cite{benaim1999dynamics,benaim2005stochastic}. Full details are provided in Appendix~\ref{app:convergence}.

%Theorem~\ref{thm:main-convergence} characterizes the limiting behavior of ESL prototype dynamics without assuming a fixed or exogenous data distribution. The ICT sets of~\eqref{eq:diff-inclusion} correspond to self-consistent regimes of the belief--policy--data feedback loop: prototype parameters that are stable under the averaged update directions induced by the invariant regimes of the fast belief process. Unlike fixed-point convergence results, the ICT characterization accommodates multiple invariant measures and non-ergodic fast dynamics, which arise naturally under endogenous interaction. Under uniqueness of the invariant measure and global asymptotic stability of the mean-field ODE, the iterates converge a.s.\ to a fixed point (Appendix~\ref{app:convergence}, Corollary~1). Prototype recovery is a separate identifiability question; the theorem concerns limiting dynamics, not ground-truth type recovery (Appendix~\ref{app:identifiability}).

Theorem~\ref{thm:main-convergence} characterizes limiting behavior without assuming a fixed data distribution. Its ICT sets represent self-consistent regimes of the coupled belief--policy--prototype dynamics and may include recurrent behavior rather than a single fixed point. Game-theoretically, each agent samples a belief-conditioned logit response to its learned prediction of the other agent's behavior. At finite inverse temperature $\lambda$, this is a quantal response; as $\lambda\to\infty$, it concentrates on utility-maximizing responses to the learned prediction, which need not equal the other agent's actual strategy. The theorem is therefore not a Nash-convergence result. Stationary fixed points may admit SCE-like and, under logit response, QRE-like interpretations under additional consistency conditions, but ESL has no general guarantee of convergence to Nash, SCE, or QRE. Under uniqueness of the invariant measure and global stability, the iterates converge almost surely to a fixed point (Appendix~\ref{app:convergence}). The theorem concerns asymptotic interaction dynamics rather than exact recovery of ground-truth latent types (Appendix~\ref{app:identifiability}); Appendix~\ref{game-theoretic} discusses the game-theoretic interpretation in more detail.

\paragraph{Theory versus practice.}
Theorem~\ref{thm:main-convergence} applies to the diminishing-step, regularized Robbins--Monro regime under Assumptions~\ref{assum:smoothness}--\ref{assum:regularity}, including the finite-batch averaging condition in Assumption~\ref{assum:fba}. The main experiments instead use constant prototype-update steps, small fixed batches ($M=2$ in the headline adaptation protocol), and no explicit $\ell_2$ prototype regularization. Constant steps permit continued prototype updates after later behavioral shifts, while small batches allow frequent updates from newly observed signals. This practical regime is not covered by the theorem: constant steps violate Assumption~\ref{assum:stepsize}, the boundedness lemma uses nonzero regularization, and fixed $M$ does not by itself establish Assumption~\ref{assum:fba}. Theorem~\ref{thm:main-convergence} therefore provides no convergence, boundedness, or finite-time guarantee for the headline PSR/DPR/LTE experiments. The diminishing-step experiment in Appendix~\ref{app:rm-diagnostic} is a theorem-facing diagnostic, not a verification of the theorem's assumptions.

%\begin{remark}[Limitation of the fixed-point corollary]
%The corollary's conditions --- uniqueness of the invariant measure and global asymptotic stability of the mean-field ODE --- are not verified for specific game classes here. The corollary is therefore a conditional result: it identifies what additional structure would yield point convergence, not a claim that ESL converges to a fixed point in the main experiments.
%\end{remark}

% \begin{remark}[Theory and practice regimes]
% The main experiments use a constant prototype step size, which prioritizes responsiveness to behavioral shifts in finite-horizon non-stationary settings. Theorem~\ref{thm:main-convergence} applies to the diminishing-step regime and should not be read as a convergence guarantee for the constant-step experiments. The diagnostic in Appendix~\ref{app:rm-diagnostic} operates in the theoretical regime and shows the qualitative signatures predicted by the theorem: bounded prototype norms, decreasing update magnitudes, and stabilizing late-window distributions. 
% \end{remark}

\section{Experiments}
\label{sec:experiments}

\paragraph{Experimental philosophy and comparator selection.}
We evaluate one focal learner against three switching opponent identities ($N=4$) on twelve fixed $3\times3$ games (four each cooperative, coordination, and adversarial). We use $A=K=3$, $T=1000$, $H=100$, switches every 200 rounds plus hazard $0.01$, and ten seeds per game ($n=120$ matched cells). Methods load identical full-precision benchmark JSONs without resampling; construction and per-game results are in Appendix~\ref{app:per-game}.

We compare ESL with Online EM, Fixed-$K$ Bayesian filtering, and switch-aware fixed-prototype Bayesian filtering (SA-FP), which adds a latent Markov prediction step but does not learn prototypes.
For the headline evaluation, let $\mathcal I_H$ contain the sampled
opponent encounters following a switch whose elapsed global-round age
is $0\le\tau\le H$ (including the switch round, $\tau=0$).
We report summed encounter-level gaps, not per-switch averages,
and an alignment-based latent tracking diagnostic where evaluated:

\begin{itemize}[leftmargin=*, itemsep=2.5pt, topsep=2pt, parsep=0pt]
     \item \textbf{Post-switch oracle regret} $\mathrm{PSR}_m(H) = \sum_{t\in\mathcal I_H}\bigl[u^\star(t)-u_m(t)\bigr],$ where $u_m(t)$ is the focal learner's realized payoff and $u^\star(t)$ is the expected payoff of a myopic best response to the true opponent-action distribution at that encounter. Lower values indicate better local post-switch adaptation. PSR is a finite-window, switch-conditioned measure rather than external or swap regret: it compares the learner's realized payoff with a myopic oracle response to the true post-switch behavior. Low PSR therefore implies neither a standard no-regret property nor an equilibrium-convergence guarantee. This is our primary adaptation metric.

     \item \textbf{Decision-relevant prediction regret} $\mathrm{DPR}_m(H) = \sum_{t\in\mathcal I_H}\bigl[\ell_m(t)-\ell^\star(t)\bigr],$ where $\ell_m(t)$ is the cross-entropy loss assigned by method $m$ to the observed opponent action, and $\ell^\star(t)$ is the loss under the true post-switch distribution.

     \item \textbf{Latent tracking error (LTE)} $\mathrm{LTE}_m(t)=-\log b_m(t)[\sigma_t(z_t^\star)]$ uses an evaluation-only Hungarian matching $\sigma_t$ from true regimes to current prototype indices. It is averaged over post-switch encounters with $0\le\tau<H$, not summed like PSR/DPR (Appendix~\ref{app:lte-definition}).
\end{itemize}
 
% \textbf{Baselines.} \textbf{PPO} \cite{schulman2017proximal} is a model-free policy-gradient learner with no opponent modeling.\textbf{ Fictitious Play (FP) }\cite{brown1951iterative} maintains per-opponent empirical action counts and best-responds to the resulting distribution. \textbf{Simple Opponent Model (SOM) }\cite{he2016opponent} predicts opponent actions from recent per-opponent history with no trainable parameters. \textbf{Fixed-K Bayesian} maintains beliefs over a fixed shared prototype library without online prototype learning. \textbf{Online EM} \cite{cappe2009line, dempster1977maximum} performs latent clustering under an i.i.d. mixture assumption, treating observations as exogenous. Each baseline removes a distinct component of ESL: PPO removes modeling, FP removes shared structure, SOM removes parameter sharing, Fixed-K removes prototype learning, and EM removes endogeneity.

\subsection{Adaptation Under Behavioral Shifts}
On the fixed twelve-game benchmark, ESL has the lowest aggregate PSR among the four online latent-adaptation methods: $162.97$ versus $187.1$ (Online EM), $203.6$ (Fixed-$K$), and $207.8$ (SA-FP). Its point estimate is lowest among all four on 8/12 games and below SA-FP on 11/12; neither count implies per-game significance (Appendix~\ref{app:per-game}).

Across 120 matched game--seed cells, SA-FP has higher mean PSR than ESL by $44.82$ (paired 95\% CI $[34.06,55.58]$, $t(119)=8.25$, $p=2.50\times10^{-13}$, $d_z=0.75$). The methods share games, seeds, opponent schedules, evaluation windows, uniform initial beliefs, and $\lambda=2$, but differ in prototype initialization and learning rules (Appendix~\ref{app:hyperparameters}). Because their policies may produce different actions and subsequent observations, this compares complete methods rather than isolating the effect of prototype updating.

In a matched-initialization control, both start from the same random prototype dictionary; ESL updates it and SA-FP keeps it fixed while predicting switches. The SA-FP--ESL gap remains $49.22$ (95\% CI $[37.71,60.73]$, $n=120$). Different exogenous random streams prevent attribution of the change in gap to initialization alone. SA-FP has the lowest DPR, while ESL has lower aggregate PSR; neither result establishes strategic co-learning or pretrained-system superiority.

\begin{figure}[t]
\centering
\includegraphics[width=\linewidth]{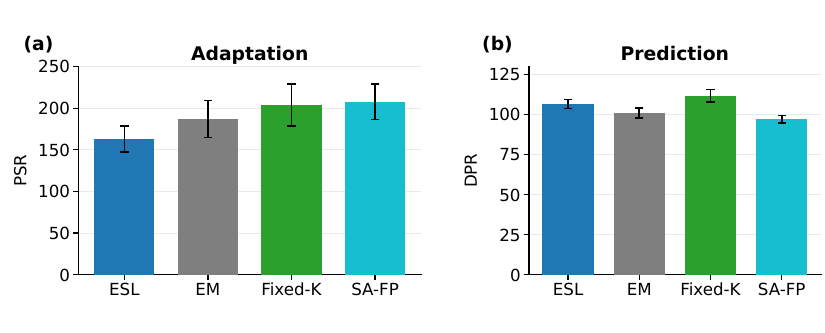}
\caption{
\textbf{Adaptation versus prediction on the twelve-game $3\times3$ benchmark.}
(a) ESL has the lowest mean post-switch oracle regret (PSR).
(b) SA-FP has the lowest mean decision-relevant prediction regret (DPR).
Error bars denote Student-$t$ 95\% confidence intervals across 120 game--seed cells
(12 games $\times$ 10 seeds). Lower values are better for both metrics;
the methods with the lowest PSR and DPR differ in this benchmark.
}
\label{fig:dynamic-adaptation}
\end{figure}

\begin{table}[H]
\centering
\caption{Mean PSR on the clean twelve-game $3\times3$ benchmark
($12$ games $\times$ $10$ seeds, lower is better). Values are rounded
to one decimal for display.}
\label{tab:dynamic-adaptation}
\begin{tabular}{lr}
\toprule
Method & PSR $\downarrow$ \\
\midrule
ESL & \textbf{163.0} \\
Online EM & 187.1 \\
Fixed-$K$ & 203.6 \\
SA-FP & 207.8 \\
\bottomrule
\end{tabular}
\end{table}

In a separately tuned, held-out evaluation, ESL has lower PSR than OMD/MWU ($86.73$ versus $116.67$; paired OMD--ESL difference $29.94$, 95\% CI $[13.79,48.07]$, $n=84$). Both methods were tuned on common development seeds before evaluation on untouched seeds, unlike the fixed-$\lambda=2$ headline comparison (Appendix~\ref{app:omd-protocol}).

\subsection{Mechanistic Analysis}

\begin{figure}[t]
\centering
\includegraphics[width=\linewidth]{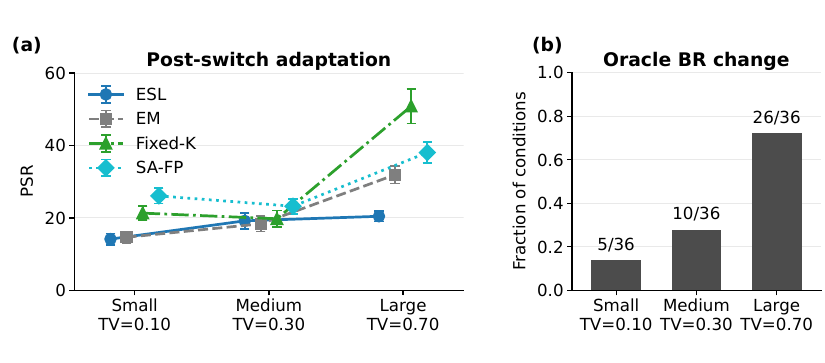}
\caption{
\textbf{Adaptation under calibrated behavioral shifts.}
On the twelve-game $3\times3$ benchmark, opponent-action distributions
undergo small, medium, or large shifts (total-variation (TV)
distances $0.10$, $0.30$, and $0.70$, respectively).
(a) Mean post-switch regret (PSR; lower is better) for four methods;
error bars are cell-level Student-$t$ 95\% confidence intervals across
360 game--rotation--seed cells per setting.
(b) Fractions of the 36 game--rotation conditions in which the
oracle best response changes; counts are deterministic for the
constructed shifts.
ESL has its clearest advantage under large shifts; Online EM has the
lowest mean PSR under medium shifts.
}
\label{fig:mechanistic}
\end{figure}

We evaluate calibrated exogenous behavioral shifts on the twelve games (Figure~\ref{fig:mechanistic}; Appendix~\ref{app:shift-difficulty}). A single shift at $t=100$ is followed by a 100-round PSR window, so regret magnitudes differ from the headline protocol. The small, medium, and large shifts have total-variation (TV) distances of $0.10$, $0.30$, and $0.70$, with Jensen--Shannon divergences of $0.010$, $0.052$, and $0.310$. They change the oracle best response in 5, 10, and 26 of 36 game--rotation conditions, respectively. This separates the magnitude of distributional change from its decision-relevant effect. ESL's advantage is clearest for large shifts, whereas Online EM has the lowest medium-shift mean. This exogenous-shift diagnostic neither isolates a causal mechanism nor directly tests Proposition~\ref{prop:likelihood-mismatch}.

\subsection{Ablation}
\begin{figure}[H]
\centering
\includegraphics[width=\linewidth]{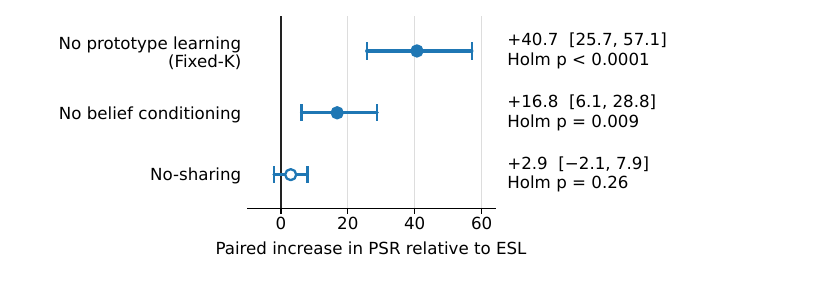}
\caption{
\textbf{Matched component ablation on twelve $3\times3$ games.}
Points show paired increases in post-switch regret (PSR) relative to
full ESL; positive values indicate worse adaptation after removal.
Horizontal bars are paired percentile-bootstrap 95\% confidence
intervals over 120 matched game--seed cells ($B=10{,}000$).
Holm-adjusted $p$-values include a fourth comparison with Online EM,
a separate baseline rather than a component ablation.
The no-sharing interval includes zero.
}
\label{fig:ablation}
\end{figure}

On the matched twelve-game protocol ($n=120$), full ESL has PSR $162.97$, DPR $106.67$, and LTE $3.23$. Removing prototype learning raises paired PSR by $40.7$ (95\% CI $[25.7,57.1]$, Holm $p<0.0001$); removing belief conditioning raises it by $16.8$ (CI $[6.1,28.8]$, Holm $p=0.009$). No-sharing's increase is $2.9$ (CI includes zero; Holm $p=0.26$). Figure~\ref{fig:ablation} and Table~\ref{tab:ablation} report paired component-removal contrasts against the same ESL reference, not separate independent estimates of ESL performance. These single-removal comparisons do not estimate interactions among components. The No-Belief interval includes zero under game-clustered uncertainty, while Fixed-$K$ uses a separately initialized dictionary; Appendix~\ref{app:hyperparameters} documents the matched-initialization control. The no-sharing interval does not establish an independent benefit of sharing.
\begin{table}[H]
\centering
\small
\caption{Matched twelve-game $3\times3$ component ablation ($n=120$ paired game--seed cells). Positive paired PSR differences favor ESL. CIs are paired percentile-bootstrap 95\% intervals ($B=10{,}000$); Holm adjustment covers four tests, including the separate Online EM comparison.}
\label{tab:ablation}
\begin{tabular}{lccc}
\toprule
\textbf{Variant} & \textbf{PSR increase vs ESL} & \textbf{Paired 95\% CI} & \textbf{$p_{\mathrm{Holm}}$} \\
\midrule
No-sharing & $+2.9$ & $[-2.1,7.9]$ & $0.26$ \\
No prototype learning (Fixed-$K$) & $+40.7$ & $[25.7,57.1]$ & $<0.0001$ \\
No belief conditioning & $+16.8$ & $[6.1,28.8]$ & $0.009$ \\
\bottomrule
\end{tabular}
\end{table}

% Full ablation figures are provided in Appendix~\ref{app:} as diagnostics.
% \paragraph{Summary.}
% Taken together, these results isolate the contribution of each component and show that adaptation benefits from the joint presence of shared structure, belief conditioning, and endogenous learning dynamics. ESL is the only method that simultaneously achieves the lowest PSR and DPR, 
% while methods that excel at latent identification (e.g., EM) fail to translate 
% this into improved adaptation.

\paragraph{Diminishing-step diagnostic.}
A separate finite-horizon pilot uses diminishing prototype steps, nonzero regularization, and growing batches on three two-action games with ten seeds each and 80 slow updates (Appendix~\ref{app:rm-diagnostic}). The observed parameter trajectories neither verify the fast-process averaging assumption nor establish the asymptotic conclusions of Theorem~\ref{thm:main-convergence}. Representational and runtime considerations are detailed in Appendices~\ref{app:complexity} and~\ref{app:runtime}.

\section{Conclusion}
ESL couples interaction-specific Bayesian beliefs with slowly learned shared behavioral prototypes under policy-dependent observations. We give a conditional stochastic-approximation characterization of the resulting closed-loop dynamics. On twelve synthetic $3\times3$ games, ESL achieves the lowest aggregate post-switch regret among the evaluated online latent-adaptation methods, although SA-FP achieves lower prediction regret. These findings distinguish decision-relevant adaptation from predictive fit; they do not establish convergence or performance guarantees for sequential-state or adversarial co-learning.

\label{sec:limitations}
\paragraph{Limitations.}
Our evaluation remains restricted to structured synthetic repeated normal-form
games with a focal learner facing exogenous switching opponents; it does not
validate complex sequential environments such as Overcooked, mixed adaptive
co-learning populations, or adversarial non-exploitability. The categorical
prototype implementation does not directly extend to high-dimensional
sequential state spaces, although state-conditioned prototype policies are a
possible untested direction. The opponent-count and population-recovery
experiments reported in Appendix~\ref{app:population-protocols} assess
tracking across identities and latent recovery, not scalability to
sequential high-dimensional state spaces. SA-FP uses frozen random emissions; stronger
comparators include learned-emission/transition HMMs, changepoint methods, and
resource-matched pretrained in-context systems such as OEOM and OMIS. The
held-out OMD experiment has limited tuning and does not compare strategic
ESL--OMD co-learning. Theorem~\ref{thm:main-convergence} is conditional on
regularity and finite-batch averaging assumptions, including an unverified
joint-process averaging condition, and applies to a diminishing-step,
regularized Robbins--Monro regime, not the constant-step, unregularized,
$M=2$ empirical implementation. Finally, $K$ is fixed within each run;
adaptive prototype counts and richer continuous, hierarchical, or fully
online representations remain future work.

\paragraph{Potential implications.}
The distinction may matter wherever predictions affect future data, including human--AI interaction and adaptive recommendations. Testing this possibility in sequential and partially observable settings remains future work.

\clearpage
\begin{ack}
J.S. is partially supported by the Infosys Center for AI and the Center
for Design and New Media (a TCS Foundation Initiative supported by
Tata Consultancy Services) at IIIT-Delhi, India.

\textbf{Competing interests.} The authors declare no competing financial
interests relevant to this work.
\end{ack}

\bibliographystyle{plain}
\bibliography{references.bib}
%%%%%%%%%%%%%%%%%%%%%%%%%%%%%%%%%%%%%%%%%%%%%%%%%%%%%%%%%%%%

\newpage
\appendix
\startcontents[eslappendix]
\section*{Table of Contents for the Appendix}
\begingroup
\small
\printcontents[eslappendix]{}{1}[2]{}
\endgroup
\clearpage
\phantomsection
\section*{Notation Guide}
\addcontentsline{toc}{section}{Notation Guide}
\noindent\textit{Quick reference for the interaction model, ESL updates, conditional theory,
and evaluation. Detailed definitions and assumptions appear in the main text and Appendix~\ref{app:notation}.}
\vspace{0.8em}

\begingroup
\normalsize
\renewcommand{\arraystretch}{1.5}
\noindent\begin{tabular}{@{}p{0.37\linewidth}p{0.59\linewidth}@{}}
\toprule
\textbf{Notation} & \textbf{Meaning} \\
\midrule
\multicolumn{2}{@{}l}{\textbf{Interaction and representation}}\\[2pt]
$N,\mathcal A,\mathcal Y,K$ & Agents, finite action set, signal space, learned prototypes \\
$t,m,M$ & Global interaction round, slow prototype-update index, observed records per batch \\
$\mathcal E_t,W_{ij,t},s_{ij,t}$ & Directed interaction set, observation indicator, received signal \\
$\theta_k,\Theta_t,\Theta_m$ & Prototype-$k$ parameters and the shared parameter collection (round/slow-update indexed) \\
$\pi_k(a\mid\theta_k),L_k(s\mid\theta_k)$ & Prototype action distribution and induced signal likelihood \\
$b_{i\to j,t},B_t$ & Agent $i$'s belief over $j$'s prototypes; collection of all pairwise beliefs \\
$\Delta_K^\delta$ & Truncated probability simplex constraining each belief \\
$\lambda,\widehat\pi_{j\mid i,t}$ & Logit-response inverse temperature and predicted opponent-action distribution \\
$\mathcal D_m,\gamma_m,\eta_{\mathrm{reg}}$ & Observed-record buffer, slow update step size, regularization coefficient \\
\addlinespace[4pt]
\multicolumn{2}{@{}l}{\textbf{Conditional stochastic-approximation theory}}\\[2pt]
$P_\Theta,Q_\Theta$ & Frozen-prototype belief transition kernel and next observed-record law (in the conditional model) \\
$H,\widehat H$ & Stacked update direction and its batch empirical average \\
$\mathcal I(\Theta),\mathcal G(\Theta)$ & Invariant-measure set and corresponding set-valued averaged mean field \\
\addlinespace[4pt]
\multicolumn{2}{@{}l}{\textbf{Evaluation metrics}}\\[2pt]
PSR & Post-switch oracle regret: payoff gap relative to a myopic best-response oracle \\
DPR & Decision-relevant prediction regret: excess cross-entropy relative to the true opponent distribution \\
LTE & Latent tracking error: evaluation-only, prototype-aligned belief diagnostic \\
\bottomrule
\end{tabular}
\endgroup
\clearpage

\section{Appendix: Convergence Analysis under Controlled Markov Noise}
\label{app:convergence}

This appendix establishes Theorem~\ref{thm:main-convergence} using stochastic approximation with controlled Markov noise \cite{borkar2006stochastic,borkar2008stochastic} and asymptotic pseudotrajectory analysis \cite{benaim1999dynamics,benaim2005stochastic}. The analysis proceeds under Assumptions~\ref{assum:smoothness},~\ref{assum:stepsize}, ~\ref{assum:moment}, \ref{assum:fba} and~\ref{assum:regularity}, together with the boundedness result of Lemma~\ref{lem:boundedness}.

\subsection{Notation and Two-Timescale Setup}
\label{app:notation}

\paragraph{Index convention.}
Let $t$ index global interaction rounds, with potentially multiple sequential encounters and masked observations per round, and let $m$ index slow prototype updates. Algorithm~\ref{algo} advances $m$ after collecting $M$ \emph{observed} interaction records, so a batch need not span $M$ rounds or $M$ encounters. In the conditional theory below, steps of $P_\Theta$ refer instead to successive observed-record transitions of a homogeneous frozen-parameter fast process. This record-indexed kernel is an additional modeling condition and need not coincide with the round-indexed transition law. For $\Theta = (\theta_1,\ldots,\theta_K) \in \mathbb{R}^{K \times |\mathcal{A}|}$ we write
\[
  \|\Theta\|
  := \|\Theta\|_F
  = \Bigl(\sum_{k=1}^K \|\theta_k\|_2^2\Bigr)^{1/2};
\]
all norms on prototype collections refer to this Frobenius norm.

\paragraph{Quasi-static approximation and mean-field map.} For the conditional theory, take the observed-record-indexed fast
belief state $B$ to evolve under a homogeneous Markov kernel $P_\Theta$
on the compact space $\mathcal B=(\Delta_K^\delta)^{N(N-1)}$
when $\Theta$ is frozen. This representation is assumed for the
fast-process model; it is not established for every empirical
interaction and switching schedule.

Under this quasi-static approximation, the limiting behavior of the
slow-timescale iterates is governed by averaged update directions
induced by the invariant regimes of the fast process. These directions
are encoded in the set-valued mean-field map
\begin{equation}
  \mathcal{G}(\Theta)
  = \bigl\{g_\mu(\Theta) : \mu \in \mathcal{I}(\Theta)\bigr\},
  \label{eq:mean-field-app}
\end{equation}
where $\mathcal{I}(\Theta)$ denotes the set of invariant probability
measures of the fast belief dynamics under frozen $\Theta$, and
\[
  g_\mu(\Theta)
  = \int_{\mathcal{B}}
    \mathbb{E}_{(i,j,s) \sim Q_\Theta(B,\cdot)}
    \bigl[H(B,i,j,s;\Theta)\bigr]\,\mu(dB),
\]
Here $H$ is the stacked vector from \eqref{eq:drift}; $Q_\Theta(B,\cdot)$ is the conditional law of the next observed record $(i,j,s)$ given the pre-record belief state $B$, while $P_\Theta$ is the transition kernel of the belief state itself. The map $\mu \mapsto
g_\mu(\Theta)$ is linear in $\mu$; convexity of $\mathcal{G}(\Theta)$
follows because $\mathcal{I}(\Theta)$ is convex (convex combinations of
invariant measures are invariant) and the image of a convex set under a
linear map is convex.

Since the fast process has no auxiliary control variable beyond the
frozen parameter $\Theta$, the relevant ergodic occupation measures
reduce to invariant probability measures in $\mathcal{I}(\Theta)$
\cite{borkar2006stochastic}. This characterization underlies the
differential inclusion in Theorem~\ref{thm:main-convergence}.

\paragraph{SA recursion.}
The prototype update induced by batch $m$ can be written compactly as
\begin{equation}
  \Theta_{m+1} = \Theta_m + \gamma_m\,\widehat{H}(\mathcal{D}_m,\Theta_m),
  \label{eq:sa-recursion-app}
\end{equation}
where $\widehat{H}$ is the empirical average of the stacked vector $H$ over the batch.
The randomness in \eqref{eq:sa-recursion-app} is induced by the fast
belief process rather than an exogenous i.i.d.\ source, placing the
recursion in the controlled Markov noise framework of
\cite{borkar2006stochastic}.

\subsection{Verification of Assumptions}

\paragraph{Verification of Assumption~\ref{assum:moment}.}
The martingale difference sequence \(\{M_{m+1}\}\) arises from centered empirical averages of belief-weighted score terms together with a linear regularization term. Since the action space is finite, beliefs lie in the compact simplex \(\Delta_K^\delta\), and the score function \(\nabla_{\theta_k}\log L_k\) is uniformly bounded, it follows that the conditional second moment of \(M_{m+1}\) grows at most linearly in \(1+\|\Theta_m\|^2\). Thus, Assumption~\ref{assum:moment} holds.

\paragraph{A sufficient condition for Assumption~\ref{assum:fba}
(Finite-Batch Averaging).}

Assumption~\ref{assum:fba} is not implied by the base ESL construction. We instead give a stronger sufficient condition under which its required window-sum property follows. For this sufficient-condition argument, suppose the fast belief
process indexed by observed records admits the homogeneous frozen kernel
$P_{\Theta_m}$ on $\mathcal B=(\Delta_K^\delta)^{N(N-1)}$ within batch $m$.
The round-indexed kernel does not automatically provide this record-indexed property. Write
$h(B;\Theta):=\mathbb E_{(i,j,s)\sim Q_\Theta(B,\cdot)}[H(B,i,j,s;\Theta)]$,
so that $g_\mu(\Theta)=\int h(\cdot;\Theta)\,d\mu$ for
$\mu\in\mathcal I(\Theta)$ and $\mathcal G(\Theta)=\{g_\mu(\Theta):\mu\in\mathcal I(\Theta)\}$
as in Section~\ref{app:notation}, and let $L_H$ denote the Lipschitz constant of
$h(\cdot;\Theta_m)$ on $\mathcal B$ (Assumption~\ref{assum:smoothness}).

\emph{Law-level geometric convergence.}
For a probability law $\nu$ on $\mathcal B$, define its distance to the invariant-measure set by
\[
d_{W_1}\bigl(\nu,\mathcal I(\Theta)\bigr)
:=
\inf_{\mu\in\mathcal I(\Theta)} W_1(\nu,\mu).
\]

\begin{assumption}[Uniform law-level geometric convergence]
\label{assum:set-contraction}
There exist constants $C<\infty$ and $\rho\in[0,1)$, uniform over the almost-sure compact parameter region of Lemma~\ref{lem:boundedness}, such that for every such $\Theta$, every $B\in\mathcal B$, and every $t\ge0$,
\[
d_{W_1}\bigl(P_\Theta^t(B,\cdot),\mathcal I(\Theta)\bigr)
\le C\rho^t.
\]
\end{assumption}

This is a stronger sufficient condition for Assumption~\ref{assum:fba}, not a property implied by the base ESL construction. It concerns convergence of the transition law toward the invariant-measure set; it does not require individual realized belief states to approach an invariant distribution. We do not establish this condition for the complete joint ESL belief process.

\emph{Batch-averaged bias under law-level mixing.}
Assumption~\ref{assum:set-contraction} and the $L_H$-Lipschitz property of $h$ imply
\[
  d\Bigl(\int h(\cdot;\Theta_m)\,dP_{\Theta_m}^t(B,\cdot),\ \mathcal G(\Theta_m)\Bigr)
  \le L_H C\rho^t.
\]
Since $\mathcal G(\Theta_m)$ is convex (Section~\ref{app:notation}), the distance to $\mathcal G(\Theta_m)$ is convex. Applying Jensen's inequality to the batch average therefore gives
\begin{equation}
\begin{aligned}
  d\Bigl(\mathbb E[\widehat H_m\mid\mathcal F_m],\ \mathcal G(\Theta_m)\Bigr)
  &\le \frac{L_H C}{M_m}\sum_{t=0}^{M_m-1}\rho^t \\
  &\le \frac{L_H C}{M_m(1-\rho)}.
\end{aligned}
\label{eq:no-uniqueness-bound}
\end{equation}

\emph{Measurable selection and the bias rate.}
Take $\xi_m := \Pi_{\mathcal G(\Theta_m)}\bigl(\mathbb E[\widehat H_m\mid\mathcal F_m]\bigr)$,
the Euclidean projection onto the compact convex set $\mathcal G(\Theta_m)$.
This projection is unique, and $\mathcal F_m$-measurable since
$\Theta\mapsto\mathcal G(\Theta)$ is upper hemicontinuous with compact
convex values (Section~\ref{app:notation} and the verification of
Assumption~\ref{assum:regularity} below) and $\Theta_m$ is
$\mathcal F_m$-measurable. By construction $\xi_m\in\mathcal G(\Theta_m)$,
and by \eqref{eq:no-uniqueness-bound},
\[
  \|\beta_m\|
  = \bigl\|\mathbb E[\widehat H_m\mid\mathcal F_m]-\xi_m\bigr\|
  = d\bigl(\mathbb E[\widehat H_m\mid\mathcal F_m],\,\mathcal G(\Theta_m)\bigr)
  \;\le\; \frac{L_HC}{M_m(1-\rho)} \;=\; O(M_m^{-1}),
\]
without requiring $\mathcal I(\Theta_m)$ to be a singleton.

\emph{Window-sum condition.}
By Lemma~\ref{lem:boundedness}, $\{\Theta_m\}$ remains almost surely in
the compact region over which $L_H,C,\rho$ in
\eqref{eq:no-uniqueness-bound} are uniform; on this probability-one
event, $\|\beta_m\|\le c_m := L_HC/(M_m(1-\rho))$ for every $m$,
deterministically along the sample path, so for $m\le n$,
\[
  \Bigl\|\sum_{j=m}^n\gamma_j\beta_j\Bigr\| \;\le\; \sum_{j=m}^n\gamma_jc_j
  \;\le\; \sum_{j=m}^\infty\gamma_jc_j.
\]
With $\gamma_m=c(m+1)^{-a}$, $a\in(1/2,1)$, and $M_m=m^{1-a+\epsilon}$ for
any $\epsilon>0$, $\gamma_mc_m = O(m^{-(1+\epsilon)})$, so
$\sum_j\gamma_jc_j<\infty$ and its tail vanishes:
$\lim_{m\to\infty}\sum_{j=m}^\infty\gamma_jc_j=0$. Taking $\sup_{n\ge m}$
above and letting $m\to\infty$,
\[
  \lim_{m\to\infty}\sup_{n\ge m}\Bigl\|\sum_{j=m}^n\gamma_j\beta_j\Bigr\|
  \;\le\; \lim_{m\to\infty}\sum_{j=m}^\infty\gamma_jc_j \;=\; 0
\]
almost surely, which is the window-sum condition of
Assumption~\ref{assum:fba}.

With fixed $M_m\equiv M$, this bound is constant in $m$ and therefore does not establish the window-sum condition in Assumption~\ref{assum:fba}. The growing-batch schedule above is one sufficient construction under the stronger contraction hypothesis; it is not claimed to hold automatically for the empirical implementation.

\paragraph{Verification of Assumption~\ref{assum:regularity}
(Regularity of the mean-field map).}

We establish the three claimed properties in turn.

\emph{Existence of invariant measures.}
For each fixed $\Theta$, the transition kernel $P_\Theta$ is Feller
(established in the kernel-continuity argument of
Step~1 below).  Since $\mathcal{B}$ is compact, the
Krylov--Bogolyubov theorem guarantees the existence of at least one
invariant probability measure, so $\mathcal{I}(\Theta) \ne \varnothing$.

\emph{Upper hemicontinuity of $\Theta \mapsto \mathcal{I}(\Theta)$.}
Let $\Theta_n \to \Theta$ and $\mu_n \in \mathcal{I}(\Theta_n)$.
Since $\mathcal{B}$ is compact metrizable, $\{\mu_n\}$ is tight;
extract a weakly convergent subsequence $\mu_{n_j}
\xrightarrow{w} \mu$ by Prokhorov's theorem.  We claim $\mu \in
\mathcal{I}(\Theta)$.

Fix any bounded continuous $f : \mathcal{B} \to \mathbb{R}$.  Because
the kernel is jointly continuous in $(B,\Theta)$ (shown below), the
map $\Theta \mapsto P_\Theta f(B)$ converges uniformly in $B$:
\[
  \sup_{B \in \mathcal{B}}
  \bigl|P_{\Theta_n}f(B) - P_\Theta f(B)\bigr|
  \;\to\; 0
  \quad\text{as } \Theta_n \to \Theta.
\]
Therefore:
\begin{align}
  \int P_\Theta f\,d\mu
  &\;=\; \lim_j \int P_{\Theta_{n_j}} f\,d\mu_{n_j}
  \label{eq:uhc-1}\\
  &\;=\; \lim_j \int f\,d\mu_{n_j}
  \label{eq:uhc-2}\\
  &\;=\; \int f\,d\mu,
  \label{eq:uhc-3}
\end{align}
where \eqref{eq:uhc-1} uses uniform kernel convergence and weak
convergence $\mu_{n_j} \xrightarrow{w} \mu$; \eqref{eq:uhc-2} uses
$\mu_{n_j}$-invariance of $P_{\Theta_{n_j}}$; and \eqref{eq:uhc-3}
uses weak convergence again.  Since $f$ was arbitrary,
$P_\Theta^*\mu = \mu$, so $\mu \in \mathcal{I}(\Theta)$.
Compactness of $\mathcal{I}(\Theta)$ follows because it is a weakly
closed subset of the weakly compact set $\mathcal{M}_1(\mathcal{B})$.

\emph{Convexity and upper semicontinuity of $\mathcal{G}(\Theta)$.}
Convexity follows from the argument in Section~\ref{app:notation}.
Upper semicontinuity of $\Theta \mapsto \mathcal{G}(\Theta)$ is
inherited from upper hemicontinuity of $\Theta \mapsto
\mathcal{I}(\Theta)$ and continuity of the averaging map
$\mu \mapsto g_\mu(\Theta)$.

\subsection{Proof of Lemma~\ref{lem:boundedness}}

\begin{proof}
Write the regularized batch update as
$\Theta_{m+1}=(1-\eta_{\mathrm{reg}}\gamma_m)\Theta_m
+\gamma_m\widehat r_m$, where $\widehat r_m$ is the empirical
belief-weighted score. Uniformly bounded categorical scores and
belief weights in $[0,1]$ give the pathwise bound
$\|\widehat r_m\|\le R:=C_0\sqrt K$ in the Frobenius norm.
By Assumption~\ref{assum:stepsize}, $\gamma_m\to0$; for all
sufficiently large $m$, $0\le\eta_{\mathrm{reg}}\gamma_m\le1$.
Hence
\[
\|\Theta_{m+1}\|
\le (1-\eta_{\mathrm{reg}}\gamma_m)\|\Theta_m\|
+\gamma_m R
\le\max\!\left\{\|\Theta_m\|,\frac{R}{\eta_{\mathrm{reg}}}\right\}.
\]
Induction bounds every tail iterate by the larger of the norm at
the start of this tail and $R/\eta_{\mathrm{reg}}$.
The finite initial prefix consists of finite updates, giving
$\sup_m\|\Theta_m\|<\infty$ almost surely. In particular,
for bounded initialization and deterministic step sizes, this
argument supplies a deterministic uniform bound.
\end{proof}

\subsection{Proof of Theorem~\ref{thm:main-convergence}}

We map the ESL recursion to the controlled-Markov stochastic approximation framework of
\cite{borkar2006stochastic}.

In the homogeneous frozen-kernel model for the observed-record clock,
the fast belief process evolves as a Markov chain on the compact state space
\[
\mathcal{B} = (\Delta_K^\delta)^{N(N-1)}.
\]
Since the transition kernel depends continuously on \(\Theta\), this defines a controlled Markov
process in the sense of \cite{borkar2006stochastic}.

\textbf{Step 1 (Noise decomposition and tracking bias).}
We first express the empirical recursion in stochastic approximation form by decomposing the batch update into its conditional expectation and fluctuation terms.

Let $\widehat{H}_m$ denote the empirical update computed from the $m$-th batch,
so that
\begin{equation}
\Theta_{m+1}
=
\Theta_m+\gamma_m \widehat{H}_m .
\label{eq:appendix-raw-update}
\end{equation}
We decompose $\widehat{H}_m$ by adding and subtracting its conditional expectation:
\begin{equation}
\widehat{H}_m
=
\xi_m
+
\underbrace{\bigl(\widehat{H}_m - \mathbb{E}[\widehat{H}_m \mid \mathcal{F}_m]\bigr)}_{M_{m+1}}
+
\underbrace{\bigl(\mathbb{E}[\widehat{H}_m \mid \mathcal{F}_m] - \xi_m\bigr)}_{\beta_m},
\label{eq:appendix-basic-decomp}
\end{equation}
where $\xi_m \in \mathcal{G}(\Theta_m)$ is a measurable selection.

Define
\[
M_{m+1}
:=
\widehat{H}_m - \mathbb{E}[\widehat{H}_m \mid \mathcal{F}_m].
\]
Then $\{M_{m+1}\}$ is a martingale difference sequence since
\[
\mathbb{E}[M_{m+1}\mid \mathcal{F}_m]=0.
\]

Similarly, define the tracking bias
\[
\beta_m
:=
\mathbb{E}[\widehat{H}_m \mid \mathcal{F}_m] - \xi_m.
\]

Substituting into \eqref{eq:appendix-raw-update} gives the stochastic approximation decomposition
\begin{equation}
\Theta_{m+1}
=
\Theta_m
+
\gamma_m\bigl(\xi_m+M_{m+1}+\beta_m\bigr),
\label{eq:appendix-sa-expanded}
\end{equation}
with drift \(\xi_m\), martingale noise \(M_{m+1}\), and tracking bias \(\beta_m\).
\medskip

The term $M_{m+1}$ captures finite-sample fluctuations around the conditional mean,
while $\beta_m$ captures the discrepancy between the finite-batch conditional drift
and the invariant-measure averaged drift.

The conditional analysis uses the homogeneous observed-record kernel
$P_{\Theta_m}$ on the compact belief space $\mathcal B$ while $\Theta_m$ is
frozen. This kernel, along with the averaging and regularity conditions,
places the modeled recursion within the controlled Markov noise framework of
\cite{borkar2006stochastic}; freezing parameters alone is insufficient.

The decomposition above is useful for interpreting the three components of the update:
\(\xi_m\) is the averaged drift, \(M_{m+1}\) is the martingale fluctuation, and
\(\beta_m\) is the discrepancy induced by finite-batch averaging. We verify the four conditions of Theorem~3.1 of
\cite{borkar2006stochastic}:

\begin{itemize}
\item \textbf{Boundedness:} the iterates \(\{\Theta_m\}\) are almost surely bounded by Lemma~\ref{lem:boundedness}.

\item \textbf{Tightness.}
The fast belief state space
$\mathcal B=(\Delta_K^\delta)^{N(N-1)}$ is compact. Hence every family of
probability measures on $\mathcal B$, including empirical occupation measures
of the frozen fast process, is tight by Prohorov's theorem. No additional
Foster--Lyapunov condition is required for the fast process.

\paragraph{Kernel continuity.}
We verify the weak continuity of $P_\Theta(B,\cdot)$ in $(B,\Theta)$.
Fix a bounded continuous test function $f:\mathcal B\to\mathbb R$. For any
sequence $(B_n,\Theta_n)\to(B,\Theta)$, consider the induced belief updates
$B'\sim P_\Theta(B,\cdot)$ and
$B_n'\sim P_{\Theta_n}(B_n,\cdot)$. The update mechanism involves:
(i) sampling interaction pairs $\mathcal E_t$,
(ii) sampling actions from policies $\pi_i(\cdot\mid b_{i\to j},\Theta)$,
(iii) observing signals $s_{ij}$, and
(iv) updating beliefs via
\[
\tilde b_{i\to j}[k]\propto b_{i\to j}[k]L_k(s\mid\theta_k),
\]
followed by projection $\Pi_{\Delta_K^\delta}$.

The interaction-pair sampling law is independent of $(B,\Theta)$. The softmax
likelihoods and logit best responses are continuous in $(B,\Theta)$, since
expected utilities are finite sums of continuous functions. The Bayesian
rescaling map is continuous because beliefs lie in $\Delta_K^\delta$ and
softmax likelihoods are strictly positive, so the normalizing denominator is
bounded away from zero on compact subsets. The Euclidean projection
$\Pi_{\Delta_K^\delta}$ is nonexpansive, hence continuous.

For the homogeneous finite-outcome kernel considered here, let $\Omega$
index the possible pair, action, and observed-signal outcomes. Write
$p_\omega(B,\Theta)$ for the probability of outcome $\omega$ and
$T_\omega(B,\Theta)$ for its resulting belief-state update. The fixed
pair/mask sampling laws and continuous finite-action policy and signal
probabilities give continuous $p_\omega$; in the observed-record clock,
conditioning presupposes a positive probability of obtaining a record.
For each fixed $\omega$, the Bayesian update and projection above make
$T_\omega$ continuous. Consequently, for bounded continuous $f$,
\[
P_\Theta f(B)=\sum_{\omega\in\Omega}
p_\omega(B,\Theta)f(T_\omega(B,\Theta))
\]
is jointly continuous in $(B,\Theta)$, as a finite sum of continuous
terms. Thus $P_\Theta(B,\cdot)$ is weakly continuous in $(B,\Theta)$,
and $P_\Theta$ is Feller for each fixed $\Theta$. This argument applies
to the assumed homogeneous observed-record model, not to arbitrary
nonstationary interaction schedules.

\item \textbf{Invariant measures and upper hemicontinuity.}
Existence of invariant measures and upper hemicontinuity of $\Theta
\mapsto \mathcal{I}(\Theta)$ are established in full in
Section~\ref{app:notation} (Verification of
Assumption~\ref{assum:regularity}).

\item \textbf{Step-size conditions:} the sequence \(\{\gamma_m\}\) satisfies Assumption~\ref{assum:stepsize}.
\end{itemize}

%Therefore, by Theorem~3.1 of \cite{borkar2006stochastic}, the interpolated process tracks the differential inclusion induced by invariant occupation measures of the frozen fast process. In this sense, the effect of incomplete averaging over finite batches is asymptotically negligible without requiring an explicit separate control argument for\(\beta_m\).

\textbf{Step 2 (Tightness of occupation measures).}
Following \cite{borkar2006stochastic}, consider the empirical occupation measures of the fast
belief process over intervals corresponding to slow-time updates. Since the fast process evolves on
the compact space
\[
\mathcal{B} = (\Delta_K^\delta)^{N(N-1)},
\]
the family of empirical occupation measures is automatically tight. Thus the tightness condition
required in the controlled-Markov stochastic approximation framework is satisfied without any
additional Foster--Lyapunov argument. Consequently, every sequence of empirical occupation
measures admits weakly convergent subsequences. By the occupation-measure characterization in
\cite[Lemma~3.1]{borkar2006stochastic}, such limit points belong to the invariant occupation-measure
set of the frozen fast process; in the present setting, since there is no auxiliary control variable
beyond the frozen parameter \(\Theta\), these reduce to invariant probability measures in
\(\mathcal{I}(\Theta)\).

\textbf{Step 3 (Asymptotic pseudotrajectory).}

\emph{Interpolated trajectory.}
Define the cumulative time
\[
T(m) := \sum_{k=0}^{m-1} \gamma_k, \quad T(0)=0,
\]
and the piecewise-linear interpolation
\[
\bar{\Theta}(t)
:=
\Theta_m
+
\frac{t - T(m)}{T(m+1)-T(m)}(\Theta_{m+1}-\Theta_m),
\quad T(m) \le t \le T(m+1).
\]
Since \(\sum_m \gamma_m = \infty\), we have \(T(m) \to \infty\).

\medskip

\emph{Noise condition.}
We verify the window-sum condition for the total noise
$\varepsilon_m = M_{m+1} + \beta_m$ in two parts.

\textbf{Martingale term.}
By Assumption~\ref{assum:moment}, the martingale differences
$M_{m+1}$ have conditional second moments bounded by
$C(1+\|\Theta_m\|^2)$. Lemma~\ref{lem:boundedness} gives an
almost-sure bound on the iterates. For each $R>0$, stop the martingale
$S_n=\sum_{j=0}^n\gamma_jM_{j+1}$ before $\|\Theta_j\|>R$;
the stopped martingale is $L^2$-bounded because
$\sum_j\gamma_j^2<\infty$, and therefore converges almost surely.
On the probability-one event of bounded iterates, a sufficiently large
$R$ is never reached, so $S_n$ itself converges almost surely to
$S_\infty$. Its noise tails consequently satisfy
\[
  \sup_{n\ge m}\left\|\sum_{j=m}^n\gamma_j M_{j+1}\right\|
  \le 2\sup_{n\ge m-1}\|S_n-S_\infty\|\longrightarrow0
  \qquad\text{almost surely}.
\]

\textbf{Tracking bias term.}
By Assumption~\ref{assum:fba} (Finite-Batch Averaging),
\[
  \lim_{m\to\infty}\sup_{n\ge m}
  \Bigl\|\sum_{j=m}^n\gamma_j\beta_j\Bigr\| = 0
  \quad\text{a.s.}
\]

\textbf{Conclusion.}
By the triangle inequality, the total noise satisfies the window-sum
condition.  Together with (i)~almost-sure boundedness of $\{\Theta_m\}$
(Lemma~\ref{lem:boundedness}), (ii)~upper semicontinuity and non-empty
convex compact values of $\mathcal{G}$
(Assumption~\ref{assum:regularity}), and (iii)~the step-size conditions
of Assumption~\ref{assum:stepsize}, all hypotheses of
\cite{benaim2005stochastic} are satisfied, and the APT conclusion follows: almost surely, on every fixed
finite forward-time window, the shifted interpolation asymptotically
tracks solution segments of $\dot\Theta\in\mathcal G(\Theta)$.
This is a trajectory-tracking statement, not a pointwise distance
between the parameter state and a set of possible velocities.

\textbf{Step 4 (Limit set characterization).}

Since \(\{\Theta_m\}\) is almost surely bounded, the interpolated trajectory
\(\bar{\Theta}(t)\) is a bounded asymptotic pseudotrajectory of the differential inclusion
\begin{equation}
\dot{\Theta} \in \mathcal{G}(\Theta).
\label{eq:appendix-di}
\end{equation}

By standard results on asymptotic pseudotrajectories for differential inclusions
\cite{benaim1999dynamics, benaim2005stochastic}, the limit set of \(\bar{\Theta}(t)\),
and hence of \(\{\Theta_m\}\), is almost surely a compact internally chain transitive (ICT)
set of the differential inclusion.

This proves Theorem~\ref{thm:main-convergence}.

\subsection{Point Convergence Under Additional Conditions}
\begin{corollary}[Conditional point convergence]
\label{cor:point-convergence}
Under the assumptions of Theorem~\ref{thm:main-convergence}, suppose
$\mathcal I(\Theta)$ is a singleton for every $\Theta$ and the resulting
mean-field ODE has a globally asymptotically stable equilibrium
$\Theta^\ast$. Then $\Theta_m\to\Theta^\ast$ almost surely.
\end{corollary}
\begin{proof}
If \(\mathcal{I}(\Theta)\) is a singleton for each \(\Theta\), then the set-valued mean-field map
\(\mathcal{G}(\Theta)\) reduces to a single-valued vector field \(g(\Theta)\), so the differential
inclusion \eqref{eq:appendix-di} collapses to the ODE
\begin{equation}
\dot{\Theta} = g(\Theta).
\label{eq:appendix-ode}
\end{equation}
By Theorem~\ref{thm:main-convergence}, the limit set $L$ of
$\{\Theta_m\}$ is almost surely an internally chain transitive set of the ODE.
Since $\Theta^\ast$ is globally asymptotically stable, its basin of attraction is
the entire state space. By the standard attractor theorem for asymptotic
pseudotrajectories, any ICT set contained in the basin of a globally attracting
equilibrium must be contained in the singleton $\{\Theta^\ast\}$. The bounded
sequence $\{\Theta_m\}$ has a nonempty compact limit set, hence
$L=\{\Theta^\ast\}$ almost surely. Therefore
\[
\Theta_m\to\Theta^\ast
\qquad\text{almost surely}.
\]
\end{proof}

\subsection{Staleness, Averaging, and Batch Size}
\label{app:staleness}

The use of buffered beliefs $b^{\mathrm{snap}}$ introduces a discrepancy between the finite-batch empirical update and an invariant-measure-averaged direction, represented by the tracking bias $\beta_m$. Assumption~\ref{assum:fba} requires this discrepancy to become negligible on the stochastic-approximation timescale. Under the stronger law-level geometric-mixing condition developed above, $\|\beta_m\|=O(M_m^{-1})$, and a suitable growing-batch schedule establishes the required window-sum condition. A fixed $M$ generally leaves a bias bound that does not vanish with $m$ and is therefore not covered by this argument.

Larger batches improve averaging but reduce responsiveness to recent behavioral shifts. The main experiments deliberately use small fixed batches and constant steps for finite-horizon adaptation; they are outside the asymptotic guarantee. Persistent interaction and strictly positive likelihoods help maintain informative updates, but they do not by themselves establish irreducibility or mixing of the complete joint belief process.

\subsection{Interpretation of ICT Limit Sets}

Theorem~\ref{thm:main-convergence} characterizes the limiting behavior of ESL through
internally chain transitive (ICT) sets of the differential inclusion \eqref{eq:appendix-di}. In the
present setting, these ICT sets correspond to self-consistent regimes of the endogenous
belief--policy--data feedback loop: for a given \(\Theta\), the fast belief dynamics induce invariant
distributions over belief states; these invariant distributions determine averaged update
directions in \(\mathcal{G}(\Theta)\); and the resulting prototype parameters remain consistent with
the data distribution generated by the induced policies.

Thus, ICT sets serve as generalized equilibrium objects for ESL under endogenous, potentially
non-ergodic dynamics. When the invariant measure is unique, the set-valued dynamics collapse
to an ordinary differential equation, in which case the ICT limit sets coincide with invariant sets
of the associated ODE, and reduce to fixed points under additional stability conditions.

\subsection{Identifiability of Learned Prototypes}
\label{app:identifiability}

The convergence result characterizes the limiting behavior of the ESL
dynamics. It does not, by itself, imply recovery of any ground-truth behavioral
types. Prototype recovery is an identifiability question.

When synthetic ground-truth prototypes $\{\theta_k^\star\}_{k=1}^K$ exist, a
local separation condition is that there exist neighborhoods
$\mathcal N_k$ of each $\theta_k^\star$ and an $\epsilon>0$ such that,
for all distinct $k,k'\in[K]$,
\[
\inf_{\theta\in\mathcal N_k,\,\theta'\in\mathcal N_{k'}}
\left\|
L_k(\cdot\mid \theta)-L_{k'}(\cdot\mid \theta')
\right\|_{\mathrm{TV}}
\ge \epsilon .
\]
This condition rules out observationally indistinguishable prototypes
locally. Recovery up to permutation would require additional identifiability,
mixing, and stability conditions that are not proved here. The empirical
adaptation and prototype-recovery diagnostics examine different outcomes;
they do not establish general recovery or show that identifiability is
unnecessary for effective adaptation. Our main theorem instead concerns
dynamical limit sets of the coupled belief--prototype process.

It is important to distinguish the ICT characterization from
classical game-theoretic equilibrium notions.  An ICT set need not be
a fixed point: it can include limit cycles or more complex recurrent
sets, all of which are consistent with the chain-recurrence structure
of the differential inclusion.  What ICT sets guarantee is
\emph{internal chain transitivity} --- any two points in the set are
connected by arbitrarily fine pseudo-orbits that remain within the
set.  This is precisely the topological notion of recurrence that
stochastic approximation iterates inherit from their limiting
flow~\cite{benaim1999dynamics}.

In the ESL setting, the ICT structure arises naturally because the
endogenous data distribution need not produce a unique invariant
regime: different interaction histories can stabilize different
belief--prototype configurations.  The theorem makes no claim that
all such configurations are equally useful for adaptation; that is an
empirical question examined in Section~\ref{sec:experiments}.

\section{Appendix: Predictive-Fit--Adaptation Mismatch}
\label{app:likelihood-adaptation-mismatch}

This appendix proves Proposition~\ref{prop:likelihood-mismatch}. Its purpose is narrow: to exhibit a setting in which predictive fit on the interaction stream and post-shift decision quality rank two behavioral representations differently. The construction uses a stylized belief-dependent policy to establish existence; it is not the experimental ESL policy.

\subsection{Closed-loop construction}

There are two latent types $z_t\in\{0,1\}$ and two actions $a_t\in\{0,1\}$. Before the switch, $z_t=0$ for $t=1,\ldots,T$; afterward, $z_t=1$. Rewards are
\[
u(a,z)=
\begin{cases}
1, & a=z,\\
-1, & a\neq z,
\end{cases}
\]
so choosing the wrong action incurs instantaneous regret $2$.

In this stylized construction, the signal space is $\mathcal Y=\{\bot,0,1\}$, with
\[
s_t=
\begin{cases}
\bot, & a_t=0,\\
z_t, & a_t=1.
\end{cases}
\]
Thus $a=0$ yields no information, whereas $a=1$ is diagnostic.

Fix $0<\varepsilon<\frac12$. For an admissible representation
\[
\Theta=(\theta_0,\theta_1)\in[\varepsilon,1-\varepsilon]^2,
\]
define
\[
P_\Theta(s=0\mid a=1,k)=\theta_k,
\qquad
P_\Theta(s=1\mid a=1,k)=1-\theta_k,
\]
and $P_\Theta(\bot\mid a=0,k)=1$.

Beliefs lie in
\[
\Delta_2^\delta
=
\{b\in\Delta_2:b(0),b(1)\ge\delta\},
\qquad
0<\delta<\frac12,
\]
and each Bayesian update is followed by Euclidean projection onto $\Delta_2^\delta$, as in the main model.

We compare
\[
\Theta^{\mathrm L}=(1-\varepsilon,1-\varepsilon)
\qquad\text{and}\qquad
\Theta^{\mathrm A}=(1-\varepsilon,\varepsilon),
\]
where $\Theta^{\mathrm L}$ collapses the two latent components and $\Theta^{\mathrm A}$ separates them. The initial belief is asymmetric,
\[
b_1(0)=\frac12+\eta,
\qquad
b_1(1)=\frac12-\eta,
\qquad
0<\eta<\frac12-\delta.
\]
Under the reward above, action $0$ uniquely maximizes expected payoff at $b_1$.

Let
\[
m(b):=|b(0)-b(1)|,
\qquad
\rho(b)=\widehat\rho(m(b)),
\]
where
\[
\widehat\rho:[0,1-2\delta]
\to
[\rho_{\min},\rho_{\max}]
\subset(0,1)
\]
is strictly decreasing and $\rho_{\min}>0$. The policy is
\[
\mathbb P(a_t=1\mid b_t)
=
\begin{cases}
\rho(b_t), & b_t(0)\ge b_t(1),\\
1, & b_t(1)>b_t(0).
\end{cases}
\]
Thus, while action $0$ is greedy, action $1$ is used as a diagnostic action; once action $1$ becomes greedy, it is selected surely. Because different representations can induce different belief trajectories, their information-acquisition rates can also differ.

For each representation $\Theta$, let
\[
\tau_T^\Theta
=
\{(b_t^\Theta,a_t^\Theta,s_t^\Theta)\}_{t=1}^{T}
\]
denote its induced pre-switch trajectory, and define its closed-loop predictive log-score
\[
\mathcal L_T^{\mathrm{cl}}(\Theta)
:=
\sum_{t=1}^{T}
\log\!\left[
\sum_{k=0}^1
b_t^\Theta(k)\,
P_\Theta(s_t^\Theta\mid a_t^\Theta,k)
\right].
\]
The representations are evaluated on their own induced trajectories: their beliefs, actions, and evidence streams can differ. For the probabilistic comparison below, the two induced systems are defined on a common product probability space with independent policy randomization.

\subsection{Predictive-score and adaptation separation}

Let
\[
q:=(1-\delta)(1-\varepsilon)+\delta\varepsilon
<1-\varepsilon,
\qquad
r:=\frac{|\log q|}{|\log(1-\varepsilon)|}>1,
\]
and define
\[
\rho_{\mathrm L}:=\rho(b_1)=\widehat\rho(2\eta).
\]
Assume
\[
r\rho_{\min}>\rho_{\mathrm L}.
\]
Then there exists $0<\xi<\rho_{\min}$ such that
\[
r(\rho_{\min}-\xi)>\rho_{\mathrm L}+\xi.
\]

Under $\Theta^{\mathrm L}$, both latent components assign identical likelihoods to every diagnostic signal, so Bayesian updating leaves the belief unchanged:
\[
b_t^{\mathrm L}=b_1
\qquad\forall t.
\]
Consequently,
\[
\mathbb P(a_t^{\mathrm L}=1\mid\mathcal F_{t-1}^{\mathrm L})
=
\rho_{\mathrm L}.
\]

Before the switch, every diagnostic observation is $s_t=0$, while non-diagnostic observations contribute $\log 1=0$ to the predictive score. Define
\[
I_t^\Theta:=\mathbf 1\{a_t^\Theta=1\},
\qquad
N_T^\Theta:=\sum_{t=1}^{T}I_t^\Theta.
\]
For $\Theta^{\mathrm L}$, every diagnostic observation has predictive probability $1-\varepsilon$, so
\[
\mathcal L_T^{\mathrm{cl}}(\Theta^{\mathrm L})
=
N_T^{\mathrm L}\log(1-\varepsilon).
\]

For $\Theta^{\mathrm A}$, projection guarantees $b_t^{\mathrm A}(1)\ge\delta$. Thus every pre-switch diagnostic observation has predictive probability at most
\[
b_t^{\mathrm A}(0)(1-\varepsilon)
+
b_t^{\mathrm A}(1)\varepsilon
\le q,
\]
and therefore
\[
\mathcal L_T^{\mathrm{cl}}(\Theta^{\mathrm A})
\le
N_T^{\mathrm A}\log q.
\]
Each pre-switch diagnostic observation favors type $0$ and increases the odds $b_t^{\mathrm A}(0)/b_t^{\mathrm A}(1)$; Euclidean projection onto $\Delta_2^\delta$ cannot reverse this ordering. Hence action $0$ remains greedy throughout the pre-switch period and
\[
\mathbb E[I_t^{\mathrm A}\mid\mathcal F_{t-1}^{\mathrm A}]
=
\rho(b_t^{\mathrm A})
\ge
\rho_{\min}.
\]

For $\Theta\in\{\Theta^{\mathrm L},\Theta^{\mathrm A}\}$, define the centered process
\[
M_T^\Theta
:=
\sum_{t=1}^{T}
\left(
I_t^\Theta-
\mathbb E[I_t^\Theta\mid\mathcal F_{t-1}^\Theta]
\right).
\]
Then $\{M_t^\Theta\}_{t=0}^{T}$ is a martingale whose increments have conditional range of length one. Moreover,
\[
N_T^{\mathrm L}=\rho_{\mathrm L}T+M_T^{\mathrm L},
\]
whereas
\[
N_T^{\mathrm A}
\ge
\rho_{\min}T+M_T^{\mathrm A}.
\]
Hoeffding--Azuma therefore gives
\[
\mathbb P\!\left(
N_T^{\mathrm L}>(\rho_{\mathrm L}+\xi)T
\right)
\le e^{-2\xi^2T},
\]
and
\[
\mathbb P\!\left(
N_T^{\mathrm A}<(\rho_{\min}-\xi)T
\right)
\le e^{-2\xi^2T}.
\]
A union bound implies that, with probability at least $1-2e^{-2\xi^2T}$,
\[
N_T^{\mathrm L}
\le(\rho_{\mathrm L}+\xi)T
<r(\rho_{\min}-\xi)T
\le rN_T^{\mathrm A}.
\]
Because both logarithms are negative and $|\log q|=r|\log(1-\varepsilon)|$, this yields
\[
\mathcal L_T^{\mathrm{cl}}(\Theta^{\mathrm A})
<
\mathcal L_T^{\mathrm{cl}}(\Theta^{\mathrm L}).
\]

We next compare post-switch regret. For this single-switch construction, the paper's PSR convention $\tau=0,\ldots,H$ contains $H+1$ evaluated rounds.

Under $\Theta^{\mathrm L}$, the latent belief remains $b_1$ after the switch, so action $0$ remains greedy and the correct action $1$ is selected only on diagnostic rounds. Hence
\[
\mathbb E[\mathrm{PSR}(\Theta^{\mathrm L})]
=
2(1-\rho_{\mathrm L})(H+1).
\]

Under $\Theta^{\mathrm A}$, each post-switch diagnostic observation is $s=1$. Before the belief reversal, $b^{\mathrm A}(1)\in[\delta,1/2]$, and such an observation strictly increases $b^{\mathrm A}(1)$; hence the lower projection boundary is inactive. While the upper boundary is also inactive, the posterior odds of type $1$ against type $0$ are multiplied by
\[
R:=\frac{1-\varepsilon}{\varepsilon}>1.
\]
If the upper projection boundary is reached, the odds have already exceeded one. Define
\[
n_\delta
=
\min\left\{
n\in\mathbb N:
R^n\frac{\delta}{1-\delta}>1
\right\}.
\]
Thus after at most $n_\delta$ diagnostic observations, action $1$ becomes greedy and is thereafter selected surely.

Before this reversal, every round is diagnostic with conditional probability at least $\rho_{\min}$. The number of non-diagnostic rounds required to obtain $n_\delta$ diagnostic observations is therefore stochastically dominated by the number of failures before $n_\delta$ successes in a Bernoulli process with success probability $\rho_{\min}$, whose expectation is
\[
\frac{n_\delta(1-\rho_{\min})}{\rho_{\min}}.
\]
Only those non-diagnostic rounds incur regret. Truncating at the finite $H+1$ evaluation window can only reduce this regret, so
\[
\mathbb E[\mathrm{PSR}(\Theta^{\mathrm A})]
\le
2\frac{n_\delta(1-\rho_{\min})}{\rho_{\min}}.
\]
Consequently, whenever
\[
H+1>
\frac{n_\delta(1-\rho_{\min})}
{\rho_{\min}(1-\rho_{\mathrm L})},
\]
the collapsed representation has strictly greater expected post-switch regret despite obtaining the higher pre-switch closed-loop predictive score with probability at least $1-2e^{-2\xi^2T}$. This proves Proposition~\ref{prop:likelihood-mismatch}.

\subsection{Matched belief-independent control}

The preceding mismatch does not require endogenous pre-switch sampling. Fix $\bar\rho\in(0,1)$ and draw
\[
D_t\overset{\mathrm{i.i.d.}}{\sim}\mathrm{Bernoulli}(\bar\rho),
\qquad t=1,\ldots,T,
\]
independently of both learners' beliefs. During the pre-switch period, impose the same diagnostic schedule on both representations:
\[
a_t=
\begin{cases}
1,&D_t=1,\\
0,&D_t=0.
\end{cases}
\]
After the switch, both learners return to the belief-conditioned policy above.

Let $\mathcal L_{T,\mathrm{exo}}^{\mathrm{cl}}(\Theta)$ denote the corresponding pre-switch predictive score. Both representations now receive exactly the same number
\[
N_T=\sum_{t=1}^{T}D_t
\]
of diagnostic observations. Hence
\[
\mathcal L_{T,\mathrm{exo}}^{\mathrm{cl}}(\Theta^{\mathrm L})
=
N_T\log(1-\varepsilon),
\]
whereas
\[
\mathcal L_{T,\mathrm{exo}}^{\mathrm{cl}}(\Theta^{\mathrm A})
\le
N_T\log q.
\]
Whenever $N_T>0$,
\[
\mathcal L_{T,\mathrm{exo}}^{\mathrm{cl}}(\Theta^{\mathrm A})
<
\mathcal L_{T,\mathrm{exo}}^{\mathrm{cl}}(\Theta^{\mathrm L}),
\]
so
\[
\mathbb P\!\left(
\mathcal L_{T,\mathrm{exo}}^{\mathrm{cl}}(\Theta^{\mathrm A})
<
\mathcal L_{T,\mathrm{exo}}^{\mathrm{cl}}(\Theta^{\mathrm L})
\right)
=
1-(1-\bar\rho)^T.
\]
The post-switch regret bounds remain
\[
\mathbb E[\mathrm{PSR}(\Theta^{\mathrm L})]
=
2(1-\rho_{\mathrm L})(H+1),
\]
and
\[
\mathbb E[\mathrm{PSR}(\Theta^{\mathrm A})]
\le
2\frac{n_\delta(1-\rho_{\min})}{\rho_{\min}}.
\]
Thus, under the same horizon condition above, the predictive-fit/adaptation mismatch persists even when pre-switch information acquisition is belief independent and matched across the two representations.

\paragraph{Scope.}
The construction establishes an existential mismatch between predictive fit and post-shift decision quality. The matched control shows that endogeneity is not necessary for the basic mismatch. Its role in the closed-loop construction is narrower: different representations induce different beliefs, which can change the amount of information acquired.

Neither result establishes that ESL learns or preserves $\Theta^{\mathrm A}$, nor that Online EM necessarily learns $\Theta^{\mathrm L}$. Whether ESL preserves useful behavioral separation is assessed separately through prototype-recovery experiments and matched component ablations. The construction therefore motivates the distinction between predictive fit and decision-relevant adaptation; it is not a learning guarantee for ESL.

\section{Experimental Details}
\label{sec:experimental_details}

\paragraph{LTE definition and interpretation.}\label{app:lte-definition}
At encounter $t$, $z_t^\star$ is the exogenous true behavioral regime and
$b_m(t)$ is method $m$'s belief over its own prototype indices \emph{before}
the observation update. For a method's current prototype emission
probabilities $q_{\ell,t}$ and a fixed mature reference emission catalog
$p_k^{\mathrm{ref}}$, the evaluation-only Hungarian assignment is
\[
\sigma_t\in\arg\min_{\sigma\in S_K}\sum_{k=1}^{K}
\mathrm{CE}\!\left(p_k^{\mathrm{ref}},q_{\sigma(k),t}\right).
\]
The learner does not observe the reference catalog or true regime labels.
The assignment is recomputed at every encounter, independently of the
belief mass assigned to the true regime. LTE records
$-\log b_m(t)[\sigma_t(z_t^\star)]$ using clipped, renormalized beliefs,
including encounters without a signal. Each game--seed cell averages its
encounter scores over switch ages $0\le\tau<H$; the reported LTE averages
the 120 cell means equally. This differs from the summed PSR/DPR window
$0\le\tau\le H$. LTE is a catalog-aligned diagnostic rather than guaranteed
recovery of behavioral types: the fixed reference catalog does not capture
all context-dependent emissions of the third regime. In the twelve-game
experiment, Online EM's LTE of $1.10\approx-\log(1/3)$ accompanies uniform
beliefs following prototype collapse and does not indicate accurate type
identification.

\subsection{Hyperparameters}
\label{app:hyperparameters}

The main adaptation evaluation and the model-capacity sensitivity
(Appendix~\ref{app:k-sweep}) use the same twelve committed $3\times3$
games and ten seeds per game. The main adaptation experiment in Figure~\ref{fig:dynamic-adaptation}
and the matched ablation in Figure~\ref{fig:ablation} use $T=1000$ and
inclusive post-switch age $0\le\tau\le H=100$ at sampled encounters. The separate diminishing-step
diagnostic has its own finite-horizon protocol (Appendix~\ref{app:rm-diagnostic}).

All methods within a figure are evaluated under matched task, seed, and
interaction schedules. No baseline is given access to switch labels, true future
types, or future observations. The hyperparameters below refer to the
finite-horizon implementation used to generate the reported figures.
Algorithm~\ref{algo} depicts the reported evaluation loop:
masked encounters neither update beliefs nor enter prototype batches;
updates occur after $M=2$ observed records, without a final flush.
The separate general-purpose trainer instead retains masked records
(zero gradient but counted in its $Q$-record batch denominator) and
flushes a final partial batch; it did not generate the reported results.

\paragraph{Headline opponent process and initialization.}
In the focal $1+3$ evaluation, one opponent identity is sampled uniformly
per round. Opponents start in regimes $0,1,2$, respectively; at every
positive multiple of 200 rounds each regime advances cyclically, while
on other rounds each opponent independently switches with hazard $0.01$
to a uniformly sampled different regime. At regime age $a$, the
three-action emission is $(1-r_a)\,u_3+r_a\,q_z$, where
$r_a=\min(a/80,1)$, $u_3$ is uniform, and $q_z$ assigns probability
$0.8$ to its preferred action and $0.1$ to each other action. For regimes
$0$ and $1$ the preferred actions are $0$ and $1$; regime $2$ uses the
previous focal action when available (otherwise action $2$). The
initial prototype-logit rows are seeded Gaussian draws
$\mathcal N(0,0.5^2)$ with $1$ added to entry $(k,k)$, while
opponent-specific beliefs start uniformly. The public release specifies
the exact random-number-stream offsets and evaluation-only LTE reference
catalog.

\begin{itemize}
\item \textbf{Number of prototypes.}
We use $K=3$ latent behavioral prototypes, matching the three opponent regimes
used in the type-shifting population.

\item \textbf{Inverse temperature.}
ESL, Fixed-K Bayesian, No-Sharing, No-Belief, and Online EM action selection use
a logit best response with inverse temperature $\lambda=2.0$. The separately evaluated FP and SOM
baselines use a softer response with $\lambda=1.4$.

\item \textbf{Belief projection.}
Bayesian belief updates are projected onto the interior simplex with floor
$\delta=10^{-4}$. Since $K=3$, this satisfies $\delta < 1/K$ and prevents
degenerate posterior beliefs at the simplex boundary.

\item \textbf{Prototype updates in constant-step experiments.}
The headline ESL update in Figure~\ref{fig:dynamic-adaptation}
uses batches of $M=2$ observed interaction records and a constant update coefficient
of $0.9$ to study finite-horizon tracking under behavioral changes.

\item \textbf{Matched component-ablation settings.}
The matched component analysis uses the twelve-game $3\times3$ protocol
with ten evaluation seeds per game, horizon $T=1000$, post-switch
parameter $H=100$, logit inverse temperature $\lambda=2$,
prototype update coefficient $\alpha=0.9$, and batch size $M=2$
fixed across variants. These settings correspond to the paired
statistics in Table~\ref{tab:ablation}.
The implemented No-Belief variant uses uniform mixtures both for action
selection and prototype-gradient weighting; Fixed-$K$ freezes a
separately initialized prototype dictionary, and No-Sharing uses
per-opponent dictionaries. Thus these are comparisons of implemented
variants, not isolated causal interventions. In game-clustered
uncertainty checks, the Fixed-$K$ paired difference remains positive,
whereas the No-Belief interval includes zero.

\item \textbf{Regularization and Robbins--Monro diagnostic.}
No explicit $\ell_2$ prototype regularization is applied in the
headline constant-step experiment or matched ablation
(Figures~\ref{fig:dynamic-adaptation} and~\ref{fig:ablation}).
The separate finite-horizon diagnostic in Figure~\ref{fig:rm-dynamics}
uses $\gamma_m=0.5(m+1)^{-0.9}$, growing batches
$M_m=\lceil50(m+1)^{0.2}\rceil$, and $\ell_2$ regularization
$\eta_{\mathrm{reg}}=10^{-3}$, for 80 completed slow updates on
three two-action games and ten seeds per game.
\end{itemize}

\paragraph{Connection to stochastic approximation.}
The headline adaptation and matched-ablation experiments use constant
prototype steps for finite-horizon tracking. The separate
Robbins--Monro pilot (Figure~\ref{fig:rm-dynamics}) uses diminishing
steps, larger frozen batches, and nonzero regularization.
It is a finite-horizon diagnostic, not evidence that the fast-process
averaging condition holds or that Theorem~\ref{thm:main-convergence}
applies to the headline implementation.

\paragraph{Baseline configuration.}
Each baseline uses a fixed configuration under the same task, seed, and
interaction schedule as ESL within the corresponding figure. Fixed-K Bayesian
uses fixed shared prototypes and Bayesian belief updates without prototype
learning. Online EM uses a sliding window of $W=50$ observations and does not
use switch labels or true types. FP maintains per-opponent empirical action
counts with Laplace initialization. SOM is a reduced conditional opponent model
with a 20-observation window. PPO~\cite{schulman2017proximal} uses the official PPO-PyTorch implementation
inside the shared interaction protocol, with actor and critic learning rates
$3\times 10^{-4}$ and $10^{-3}$, discount factor $0.99$, clipping parameter
$0.2$, and four policy epochs per update.

\paragraph{Switch-aware filtering and matched-initialization controls.}
The switch-aware fixed-prototype baseline (SA-FP) uses the
\texttt{fixed\_logits} library, a game-independent, non-oracle random
dictionary drawn with seed 314159 and fixed before evaluation. It is
neither fitted to benchmark outcomes nor initialized from the true
opponent types. SA-FP freezes this dictionary, predicts its latent
belief through a three-state Markov transition matrix with
$T_{kk}=0.95$ and $T_{k\ell}=0.025$ for $k\neq\ell$, and then
performs the Bayesian evidence update. The transition probabilities
were specified independently of the benchmark's scheduled switches
every 200 rounds and extra hazard of $0.01$; SA-FP has access to neither
the actual switch labels nor the true switching mechanism.

The standard $n=120$ paired comparison matches games, evaluation
seeds, horizon, opponent schedule, switch realization, post-switch
window, uniform initial beliefs, and logit policy form
($\lambda=2$). ESL instead begins with its native seeded
initialization and updates prototypes online. Hence the standard
SA-FP-minus-ESL PSR difference of $44.82$ (95\% CI
$[34.06,55.58]$) is not a one-factor causal estimate of the
benefit of prototype learning.

As a separate control, both methods start from the same
\texttt{fixed\_logits} dictionary. ESL updates that dictionary,
whereas SA-FP holds it fixed while applying Markov belief prediction.
The paired SA-FP-minus-ESL-with-fixed-initialization difference remains
$49.22$ PSR points (95\% CI $[37.71,60.73]$; paired
$t(119)=8.47$, $p=7.71\times10^{-14}$, $d_z=0.77$).
Under this matched initialization, Fixed-$K$ minus ESL-with-fixed-
initialization is $49.46$ (95\% CI $[32.27,66.64]$), while Fixed-$K$
minus SA-FP is $0.24$ (95\% CI $[-13.70,14.18]$). Thus no
significant additional PSR benefit of SA-FP's particular Markov prior
over Fixed-$K$ was detected in this control. Exogenous random-number
streams were separated in the matched-initialization rerun, so the
change from $44.82$ to $49.22$ cannot be attributed to initialization
alone. Even for matched exogenous conditions, realized opponent-action
evidence may diverge as the agents' policies affect their observations.
The additional control therefore supports persistence of ESL's
empirical advantage under shared initial prototypes, not a causal
decomposition or a guarantee against strategic co-learning opponents.

\paragraph{Reproducibility and aggregation.}
Full-mode reproduction uses seeds $0,\ldots,9$ for the twelve-game
adaptation protocol. Other diagnostics state their own seed splits.
Figures and tables use the experiment-specific aggregation and uncertainty
conventions stated in their captions or supplementary protocols.
The released code, frozen games, and numerical evidence are available in the
\href{https://github.com/jainendra2019/esl/releases/tag/neurips2026-flagship-cli}{publication-ready ESL repository}; its README distinguishes supporting plots from the manuscript figures.
\subsection{Population-Size Protocols}
\label{app:population-protocols}
We distinguish three non-interchangeable population protocols. In the
headline adaptation benchmark, one focal learner faces three persistent
opponent identities ($1+3$, $N=4$), and PSR is aggregated across twelve
committed games and ten seeds. The opponent-count sensitivity instead
tracks $J\in\{3,10,100\}$ identities ($1+J$). Switching and evaluation
are indexed by each opponent's encounter count, so per-opponent evidence,
switch frequency, and exposure to the post-switch window are matched
across $J$. The mean PSR per switch under this encounter-normalized protocol is $6.75$, $5.88$,
and $5.40$ for $J=3$, $10$, and $100$, respectively. These values belong
to the encounter-normalized protocol and must not be pooled with the
headline PSR aggregation. They examine opponent-count sensitivity, not
wall-clock, sample-complexity, or sequential-state scalability.

A distinct population-recovery diagnostic samples interaction pairs
stochastically in populations with $N=20$ and $N=100$, maintaining
interaction-specific beliefs and a population-shared prototype dictionary.
Its reported aggregate latent-recovery accuracies are $0.955$ and $0.847$,
respectively. This is a type-recovery task rather than an adaptation-PSR
evaluation, and these accuracies cannot be compared numerically with the
opponent-count PSR figures.

\subsection{Action-Selection and Update Sensitivity}
\label{app:sensitivity-protocols}
In the main twelve-game adaptation configuration, $\lambda=2$,
prototype-update coefficient $\alpha=0.9$, and batch size $M=2$ were
pre-specified, not retuned on these sensitivity results. We vary
$\alpha\in\{0.1,0.3,0.5,0.7,0.9\}$ and $M\in\{1,2,5,20\}$
one at a time on the same twelve fixed $3\times3$ games and ten seeds
per game ($n=120$ game--seed cells per setting); the
$\lambda\in\{0,0.5,2,5,10\}$ sweep follows a separate controlled
fixed-opponent-evidence protocol. In the implementation, $\alpha$
(the code's $\gamma$) scales a batch-averaged prototype-gradient
step, while $M$ counts observed opponent actions pooled globally
before that step; an incomplete terminal batch is discarded.
Unlike the controlled $\lambda$ sweep, these $\alpha$ and $M$ runs
retain interactive opponent-response dynamics, so changing one
parameter can change subsequent opponent trajectories.

\begin{table}[t]
\centering
\small
\setlength{\tabcolsep}{4pt}
\begin{tabular}{llrr}
\toprule
Sweep & Setting & PSR $\downarrow$ (mean $\pm$ SEM) & DPR $\downarrow$ \\
\midrule
$\alpha$ ($M=2$) & $0.1$ & $187.32\pm11.24$ & $106.72$ \\
 & $0.3$ & $171.61\pm9.15$ & $99.27$ \\
 & $0.5$ & $164.97\pm8.37$ & $98.50$ \\
 & $0.7$ & $162.88\pm7.79$ & $101.44$ \\
 & $0.9$ (default) & $162.97\pm7.80$ & $106.67$ \\
\midrule
$M$ ($\alpha=0.9$) & $1$ & $170.67\pm7.99$ & $141.27$ \\
 & $2$ (default) & $162.97\pm7.80$ & $106.67$ \\
 & $5$ & $170.63\pm8.96$ & $101.73$ \\
 & $20$ & $191.05\pm11.71$ & $112.86$ \\
\bottomrule
\end{tabular}
\caption{Update-coefficient and batch-size sensitivity for ESL on the
clean twelve-game $3\times3$ suite (ten seeds per game). Each sweep
changes one parameter while holding the other at its pre-specified
default. The reported uncertainties are standard errors of the mean
(SEM), \emph{not} 95\% confidence intervals.}
\label{tab:alpha-m-sensitivity}
\end{table}

PSR improves as $\alpha$ increases from $0.1$ to $0.7$ and then
plateaus: $\alpha=0.7$ and $0.9$ differ by only $0.09$ PSR points.
The smallest observed PSR across the tested batch sizes occurs at
$M=2$; DPR follows a different pattern in both sweeps
(Table~\ref{tab:alpha-m-sensitivity}). Smaller batches update more
often and larger ones delay aggregation, but no direct gradient-noise
or staleness decomposition was measured. Observation-noise sensitivity
uses the same 3-action corruption grid as the robustness protocol in
Appendix~\ref{app:robustness-protocols}, rather than a separate sweep.

The controlled focal-policy $\lambda$ sweep evaluates all five settings
on the same twelve $3\times3$ games and ten seeds per game
($n=120$ game--seed cells per setting), with one focal learner and three
opponents, $T=1000$, $H=100$, switches every 200 rounds plus
hazard $0.01$, full noiseless observation, $\alpha=0.9$, and $M=2$.
Separate random streams regenerate identical exogenous opponent actions,
switch schedules, and observation evidence for each $\lambda$; only the
focal action-sampling stream changes. Beliefs and prototypes continue to
update, using opponent evidence rather than sampled focal actions.
Consequently, DPR is exactly invariant ($100.474$) across this sweep,
while changing $\lambda$ alters the focal logit policy and its PSR
(Table~\ref{tab:focal-lambda-sensitivity}). The lower PSR at larger
$\lambda$ is a result of this controlled protocol, not a general
closed-loop co-learning claim. In particular, the $\lambda=2$ PSR of
$158.44$ here differs from the headline $162.97$ because the
experiments use different random streams, even though they share
the game suite and nominal parameters.

\begin{table}[t]
\centering
\small
\resizebox{\linewidth}{!}{%
\begin{tabular}{ccccc}
\toprule
$\lambda$ & PSR (95\% CI) & $\Delta$PSR vs.\ $\lambda=2$ (95\% paired CI) &
Action entropy & $p(\mathrm{best})$ \\
\midrule
$0$   & $517.410\ [465.099,569.721]$ & $+358.97\ [313.87,404.06]$ & $1.098612$ & $0.333333$ \\
$0.5$ & $357.933\ [328.948,386.917]$ & $+199.49\ [177.91,221.08]$ & $1.030446$ & $0.457469$ \\
$2$   & $158.440\ [142.908,173.973]$ & ---                          & $0.702508$ & $0.674042$ \\
$5$   & $94.458\ [79.013,109.903]$   & $-63.98\ [-70.52,-57.45]$  & $0.396284$ & $0.823255$ \\
$10$  & $80.947\ [64.388,97.506]$    & $-77.49\ [-85.13,-69.86]$  & $0.228029$ & $0.898461$ \\
\bottomrule
\end{tabular}%
}
\caption{Controlled focal-policy inverse-temperature sensitivity
($12$ games $\times$ $10$ seeds per setting). PSR intervals are
ordinary Student-$t$ 95\% CIs across game--seed cells; contrasts
use paired 95\% CIs against the pre-specified main $\lambda=2$.
Prediction regret (DPR) equals $100.474$ at every setting because
the opponent evidence is unchanged.}
\label{tab:focal-lambda-sensitivity}
\end{table}

\subsection{Partial Observation, Noise, and Incentive-Asymmetry Protocols}
\label{app:robustness-protocols}
The partial-observation and observation-noise experiments separately
perturb the same twelve fixed clean $3\times3$ games and switching protocol
used in the headline comparison (ten seeds per game). In the visibility
sweep, the \emph{focal learner's} signal is masked by
$W_t\sim\operatorname{Bernoulli}(p_{\mathrm{obs}})$,
$p_{\mathrm{obs}}\in\{1.0,0.8,0.6,0.4\}$; the mask is matched across
methods but does not freeze their realized opponent-action sequences.
When $W_t=0$, ESL, Fixed-$K$, and Online EM skip evidence-based updates;
SA-FP retains its Markov prediction step but makes no signal-based
Bayesian correction. Unobserved actions do not enter ESL's prototype
update batch. In the separate noise sweep, $p_{\mathrm{obs}}=1$ and
$\epsilon\in\{0,0.05,0.10,0.20\}$ controls the symmetric three-action
corruption channel
\[
P(\tilde a_t=a_t\mid a_t)=1-\epsilon,\qquad
P(\tilde a_t=a'\mid a_t)=\frac{\epsilon}{2}\quad(a'\ne a_t).
\]
Only the observed learning signal is corrupted; payoffs and regret use
the true opponent action, and the learner treats the corrupted signal
as clean. The two sweeps are not a joint mask-and-noise grid.
Table~\ref{tab:robustness-visibility-noise} reports all seven evaluated
methods; Online EM and PPO results were added in a later matched-protocol
run. For ESL, visibility PSR across $p_{\mathrm{obs}}=1,0.8,0.6,0.4$
is $162.97$, $167.69$, $173.09$, and $182.45$ (Student-$t$ 95\%
CIs $[147.52,178.42]$, $[151.15,184.23]$,
$[155.86,190.32]$, $[162.55,202.34]$). Across
$\epsilon=0,0.05,0.10,0.20$, ESL PSR is $162.97$, $164.22$,
$170.99$, and $177.04$ (95\% CIs $[147.52,178.42]$,
$[148.17,180.26]$, $[153.52,188.46]$, and
$[158.69,195.38]$). ESL has the lowest aggregate mean PSR at every
tested level, but not in every family: at $p_{\mathrm{obs}}=0.4$ on
cooperative games, Fixed-$K$ has PSR $105.82$ versus ESL $106.27$.
At $\epsilon=0.20$, a hierarchical bootstrap CI for the
Fixed-$K$--ESL gap includes zero, so aggregate rank does not imply
a robustly significant difference for that comparison.

\begin{table}[t]
\centering
\small
\resizebox{\linewidth}{!}{%
\begin{tabular}{llrrrrrrr}
\toprule
Sweep & Level & ESL & Online EM & Fixed-$K$ & SA-FP & FP & SOM & PPO \\
\midrule
Visibility & $1.0$ & \textbf{163.0} & 187.1 & 203.6 & 207.8 & 251.5 & 237.5 & 391.2 \\
 & $0.8$ & \textbf{167.7} & 189.1 & 202.8 & 212.8 & 251.3 & 243.4 & 389.3 \\
 & $0.6$ & \textbf{173.1} & 192.2 & 203.2 & 221.9 & 253.5 & 255.8 & 388.6 \\
 & $0.4$ & \textbf{182.4} & 196.2 & 205.3 & 236.1 & 254.3 & 264.2 & 387.4 \\
\midrule
Noise & $0$ & \textbf{163.0} & 187.1 & 203.6 & 207.8 & 251.5 & 237.5 & 391.2 \\
 & $0.05$ & \textbf{164.2} & 187.1 & 200.8 & 208.6 & 250.3 & 241.8 & 389.0 \\
 & $0.10$ & \textbf{171.0} & 188.7 & 199.6 & 211.3 & 253.3 & 245.4 & 386.3 \\
 & $0.20$ & \textbf{177.0} & 190.7 & 195.3 & 218.4 & 251.6 & 247.5 & 392.1 \\
\bottomrule
\end{tabular}%
}
\caption{Partial-observation and observation-noise robustness on
the twelve fixed symmetric $3\times3$ games. Entries are mean
post-switch regret (PSR; lower is better), rounded to one decimal,
across $n=120$ game--seed cells per method and setting. The two
perturbations are varied separately; the bold values identify the
smallest aggregate PSR at each level, not universal per-family wins.}
\label{tab:robustness-visibility-noise}
\end{table}

Incentive asymmetry is evaluated separately on the fixed, file-backed
\texttt{benchmark\_asym3} suite of twelve $3\times3$ general-sum
bimatrix games: three interaction families (cooperative, coordination,
and adversarial) crossed with four incentive templates
(reward preference, penalty, equilibrium conflict, and combined).
Each player's payoff matrix is independently normalized to $[-1,1]$.
Both focal objectives are evaluated by exchanging the two stored payoff
matrices \emph{without transposing their action indices}, as implemented
by the focal-role swap; this is not the conventional transpose-based
role reversal of a bimatrix game. All runs use clean observations and
ten seeds per game and focal objective ($n=240$ cells per method).
Table~\ref{tab:robustness-asymmetric} shows the key latent-adaptation
comparators; the other evaluated methods have overall PSR $179.37$
(PPO), $207.63$ (SOM), and $210.36$ (FP). ESL has the lowest
aggregate PSR in both focal objectives, with the advantage concentrated
in coordination and adversarial families. Cooperative cells are close:
SA-FP and Fixed-$K$ have slightly lower mean PSR than ESL, and the
paired SA-FP--ESL difference there ($-1.3$, 95\% CI
$[-2.9,0.3]$) includes zero. Across all game--role--seed cells,
the paired SA-FP--ESL difference is $+16.64$ (95\% CI
$[13.49,19.79]$, $n=240$). Asymmetry results are not pooled with
the symmetric headline comparison.

\begin{table}[t]
\centering
\small
\resizebox{\linewidth}{!}{%
\begin{tabular}{lrrrr}
\toprule
Asymmetric suite & ESL & Online EM & Fixed-$K$ & SA-FP \\
\midrule
Overall ($n=240$) & \textbf{155.15} & 171.37 & 171.60 & 171.78 \\
\quad 95\% CI & $[149.13,161.16]$ & $[164.01,178.72]$ & $[163.88,179.33]$ & $[164.17,179.40]$ \\
\midrule
Row focal ($n=120$) & \textbf{155.51} & 173.15 & 171.81 & 175.89 \\
Column focal ($n=120$) & \textbf{154.78} & 169.59 & 171.40 & 167.68 \\
\midrule
Cooperative ($n=80$) & 106.87 & 107.35 & 105.75 & \textbf{105.56} \\
Coordination ($n=80$) & \textbf{183.11} & 212.91 & 216.14 & 216.29 \\
Adversarial ($n=80$) & \textbf{175.46} & 193.85 & 192.92 & 193.50 \\
\bottomrule
\end{tabular}%
}
\caption{Mean PSR (lower is better) on a distinct, clean asymmetric
$3\times3$ suite of three interaction families crossed with four
incentive templates, evaluated under both implemented focal objectives.
Overall intervals are Student-$t$ 95\% CIs across game--role--seed
cells; subgroup entries are descriptive means. Bold highlights the
smallest mean in each grouping.}
\label{tab:robustness-asymmetric}
\end{table}

\subsection{Held-Out-Seed OMD Comparison}
\label{app:omd-protocol}
We include a representative online mirror descent/multiplicative-weights
(OMD/MWU) comparator for policy-level online learning in repeated games.
This is a distinct comparison from the fixed-$\lambda=2$ headline
latent-adaptation evaluation, not an additional entry in
Table~\ref{tab:dynamic-adaptation}. Each method is evaluated separately
as the focal learner against the same exogenous switching-opponent environment;
we did not evaluate a heterogeneous population in which ESL and OMD learn
against one another.

The development seeds are $\{0,1,2\}$, and the untouched confirmatory
seeds are $\{3,4,5,6,7,8,9\}$. We select one decision hyperparameter per
method using the same development seeds and mean post-switch oracle regret
(PSR) criterion, then freeze both choices before confirmatory evaluation.
For ESL, the inverse-temperature candidates are
$\lambda\in\{0,0.5,2,5,10\}$, giving $\lambda^*=10$.
For OMD/MWU, the learning-rate candidates are
$\eta\in\{0.01,0.03,0.1,0.3,1.0\}$, giving $\eta^*=1.0$.
This matched budget avoids judging the learning-rate-sensitive OMD
comparator at an arbitrary weak setting. Both use the same twelve committed
$3\times3$ games, focal $1+3$ architecture, horizon $T=1000$,
post-switch window $H=100$, switches every 200 rounds, and additional
hazard $0.01$. The seven evaluation seeds across twelve games yield
$n=84$ game--seed observations.

On these held-out cells, mean PSR is $86.73$ for tuned ESL and $116.67$ for
tuned OMD, with mean paired OMD-minus-ESL difference $29.94$ (95\% CI
$[13.79,48.07]$). These results offer a limited comparison with a
policy-level no-regret family under matched tuning and switching exogenous
opponents; they do not establish external/swap-regret guarantees,
strategic exploitability bounds, or ESL--OMD co-learning performance.

\subsection{Per-Game Breakdown}
\label{app:per-game}

\paragraph{Committed benchmark and evaluation.}
The clean benchmark contains twelve $3\times3$ normal-form games:
\texttt{coop3\_01}--\texttt{coop3\_04},
\texttt{coord3\_01}--\texttt{coord3\_04}, and
\texttt{adv3\_01}--\texttt{adv3\_04}.
All evaluated methods use the same committed game instances; the matrices are not
resampled between methods, seeds, or runs. In each game, one focal agent
interacts with three persistent opponent identities ($N=4$), with $A=K=3$,
horizon $T=1000$, and post-switch window $H=100$. Opponents undergo scheduled
type switches every 200 rounds and an additional switching hazard of $0.01$.
Each method is evaluated on ten seeds per game, giving 120 game--seed cells.
The overall mean PSR is the unweighted mean of these cells, equivalently the
mean of the twelve per-game means because each game has ten seeds.

\paragraph{Deterministic game selection.}
The documented game-generation procedure uses NumPy's \texttt{Generator} with
seed \texttt{20260728}, producing 60 candidates per family in the fixed order
cooperative, coordination, adversarial. The cooperative candidates draw
$r\sim\operatorname{Unif}(2,5)$,
$t\sim\operatorname{Unif}(r+0.5,r+3)$,
$p\sim\operatorname{Unif}(0,1.5)$,
$s\sim\operatorname{Unif}(-1,0.5)$, and a bottom-right entry from
$\operatorname{Unif}(p,p+1)$, denoted $u$, with row payoff
$U_{\rm coop}=\left[\begin{smallmatrix}r&s&s\\t&p&p\\0.8t&p&u\end{smallmatrix}\right]$
and column payoff $U_{\rm coop}^{\mathsf T}$. An unused $3\times3$
matrix of i.i.d. $\operatorname{Unif}(0,1)$ entries is also drawn for
each candidate to preserve the random-number stream. Coordination candidates draw
$r\sim\operatorname{Unif}(3,6)$,
$t\sim\operatorname{Unif}(1,r-0.5)$,
$p\sim\operatorname{Unif}(0.5,t)$,
$s\sim\operatorname{Unif}(-1,0.5)$, and
$m\sim\operatorname{Unif}(p,r)$, with identical row and column
payoffs $U_{\rm coord}=\operatorname{diag}(r,m,p)+
s(\mathbf 1\mathbf 1^{\mathsf T}-I_3)$. Adversarial candidates draw the nine row payoffs
i.i.d. from $\mathcal N(0,1)$, mean-center the resulting matrix, multiply
by $\alpha\sim\operatorname{Unif}(0.5,3)$, and set the column-player payoff
to its negative.

Diversity filtering uses unnormalized feature vectors and Euclidean distance.
Cooperative and coordination vectors contain the nine flattened row payoffs
and, respectively, the three gaps $(t-r,r-p,p-s)$ or $(r-t,t-p,m-p)$;
adversarial vectors contain only the nine flattened row payoffs. Four games
per family are selected by a greedy farthest-point filter: candidates are
considered in generation order and accepted when their minimum distance
to the already selected games is at least $1.25$. If fewer than four games
are selected, remaining candidates are added by maximizing that minimum
distance, breaking ties by earlier candidate index, with the reported
$0.625$ early-stop threshold. The twelve games are fixed, file-backed instances; the evaluation
loader reads their original full-precision JSON payoffs rather than resampling.
The files under \texttt{benchmark/} reproduce with exact NumPy array
equality (row and column payoffs, NumPy 2.2.6) from
\texttt{generate\_3x3\_suite(seed=20260728, per\_family=4)}
in \texttt{esl/res/benchmark.py}. The CLI wrapper
\texttt{scripts/generate\_res\_benchmark.py} instead adds three to
its base seed when generating the 3-action suite (default base seed
\texttt{20260725}); passing \texttt{20260728} as its CLI base seed
would generate a different suite. The original experiment CSVs record
game identifiers rather than payoff hashes; independent archived
diversity statistics corroborate the recovered instances, without
providing a byte-level hash of the payoffs in every historical run.

\begin{figure}[t]
\centering
\includegraphics[width=\linewidth]{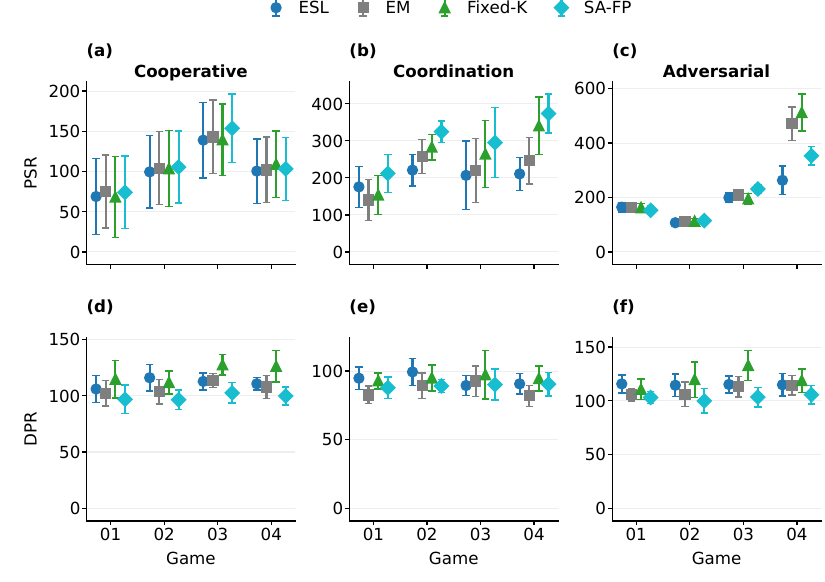}
\caption{
\textbf{Per-game adaptation and prediction on the twelve-game $3\times3$ benchmark.}
Columns show cooperative, coordination, and adversarial games (four
instances each); the top row reports post-switch regret (PSR) and
the bottom row decision-relevant prediction regret (DPR).
Points are means and error bars are ordinary Student-$t$ 95\%
confidence intervals across ten seeds per method and game.
Lower is better for both metrics. ESL has the lowest PSR point estimate
among the four methods on 8/12 games and a lower estimate than SA-FP
on 11/12; these per-game comparisons are descriptive, not
paired significance tests.
}
\label{fig:per-game}
\end{figure}

\paragraph{Per-game results.}
Table~\ref{tab:pergame-psr-ci} reports means and ordinary 95\% confidence
intervals over ten seeds per method and game. ESL has the lowest overall-method
point estimate on eight of twelve games, while its PSR point estimate is
lower than SA-FP's on eleven of twelve games. The per-game intervals are
not paired method-difference intervals and do not establish significance
of individual game comparisons. The aggregate matched ESL--SA-FP test is
reported in the main experimental comparison.

\begin{table}[H]
\centering
\scriptsize
\caption{Per-game mean PSR with ordinary 95\% CIs over seeds ($n{=}10$ per cell).}\label{tab:pergame-psr-ci}
\begin{tabular}{lcccc}
\toprule
Game & ESL & Online EM & Fixed-$K$ & SA-FP \\
\midrule
\texttt{coop3\_01}
  & $68.9$ $[21.3,116.4]$
  & $75.3$ $[29.9,120.7]$
  & $68.3$ $[18.2,118.4]$
  & $74.0$ $[29.0,119.0]$ \\
\texttt{coop3\_02}
  & $99.5$ $[54.4,144.6]$
  & $104.2$ $[58.7,149.6]$
  & $103.5$ $[56.3,150.7]$
  & $105.4$ $[60.8,150.1]$ \\
\texttt{coop3\_03}
  & $139.0$ $[91.9,186.0]$
  & $143.0$ $[97.3,188.6]$
  & $139.4$ $[94.8,183.9]$
  & $153.6$ $[111.1,196.2]$ \\
\texttt{coop3\_04}
  & $100.5$ $[60.4,140.6]$
  & $102.0$ $[61.2,142.8]$
  & $108.9$ $[67.4,150.3]$
  & $103.2$ $[64.1,142.3]$ \\
\texttt{coord3\_01}
  & $175.6$ $[120.6,230.7]$
  & $140.2$ $[84.6,195.7]$
  & $153.6$ $[100.9,206.4]$
  & $211.9$ $[160.9,262.9]$ \\
\texttt{coord3\_02}
  & $220.8$ $[178.0,263.5]$
  & $257.7$ $[211.5,303.8]$
  & $282.9$ $[248.6,317.3]$
  & $324.4$ $[294.9,353.9]$ \\
\texttt{coord3\_03}
  & $206.6$ $[113.9,299.3]$
  & $219.5$ $[132.9,306.1]$
  & $264.1$ $[173.7,354.6]$
  & $295.1$ $[200.6,389.5]$ \\
\texttt{coord3\_04}
  & $210.6$ $[166.3,254.9]$
  & $246.0$ $[183.5,308.4]$
  & $340.2$ $[262.7,417.7]$
  & $373.5$ $[321.1,425.9]$ \\
\texttt{adv3\_01}
  & $164.5$ $[147.3,181.7]$
  & $163.9$ $[147.5,180.3]$
  & $162.5$ $[146.2,178.8]$
  & $153.0$ $[137.4,168.6]$ \\
\texttt{adv3\_02}
  & $107.2$ $[97.3,117.2]$
  & $113.1$ $[103.2,123.1]$
  & $112.7$ $[101.9,123.6]$
  & $115.0$ $[104.7,125.4]$ \\
\texttt{adv3\_03}
  & $199.3$ $[182.4,216.2]$
  & $209.4$ $[189.5,229.4]$
  & $195.2$ $[176.3,214.2]$
  & $231.3$ $[214.1,248.5]$ \\
\texttt{adv3\_04}
  & $263.1$ $[210.3,315.9]$
  & $470.6$ $[409.2,531.9]$
  & $512.3$ $[444.8,579.8]$
  & $353.0$ $[319.3,386.7]$ \\
\bottomrule
\end{tabular}
\end{table}

\paragraph{Committed payoff instances (rounded display).}
For each $3\times3$ matrix below, entry $(i,j)$ gives the row-player payoff
when the row and column players choose actions $i$ and $j$.
The column payoff is the transpose of the row payoff in the cooperative
family, equals the row payoff in the coordination family, and is its negative
in the adversarial family. Numerical entries are displayed to three decimal
places; these rounded displays are insufficient for bit-exact reproduction without the original full-precision committed game instances.

\paragraph{Cooperative games.}
\begin{align*}
U_{\texttt{coop3\_01}} &=
\begin{bmatrix}
2.041 & 0.466 & 0.466 \\
5.025 & 1.369 & 1.369 \\
4.020 & 1.369 & 2.279
\end{bmatrix},
&
U_{\texttt{coop3\_02}} &=
\begin{bmatrix}
3.459 & 0.085 & 0.085 \\
4.094 & 0.208 & 0.208 \\
3.275 & 0.208 & 1.100
\end{bmatrix},
\\[1em]
U_{\texttt{coop3\_03}} &=
\begin{bmatrix}
2.609 & 0.388 & 0.388 \\
4.933 & 0.771 & 0.771 \\
3.946 & 0.771 & 1.135
\end{bmatrix},
&
U_{\texttt{coop3\_04}} &=
\begin{bmatrix}
4.454 & -0.421 & -0.421 \\
4.958 & 0.957 & 0.957 \\
3.966 & 0.957 & 1.774
\end{bmatrix}.
\end{align*}

\paragraph{Coordination games.}
\begin{align*}
U_{\texttt{coord3\_01}} &=
\begin{bmatrix}
5.978 & 0.225 & 0.225 \\
0.225 & 2.011 & 0.225 \\
0.225 & 0.225 & 1.587
\end{bmatrix},
&
U_{\texttt{coord3\_02}} &=
\begin{bmatrix}
3.011 & -0.398 & -0.398 \\
-0.398 & 1.218 & -0.398 \\
-0.398 & -0.398 & 0.853
\end{bmatrix},
\\[1em]
U_{\texttt{coord3\_03}} &=
\begin{bmatrix}
5.382 & 0.253 & 0.253 \\
0.253 & 2.564 & 0.253 \\
0.253 & 0.253 & 1.726
\end{bmatrix},
&
U_{\texttt{coord3\_04}} &=
\begin{bmatrix}
5.337 & -0.323 & -0.323 \\
-0.323 & 3.048 & -0.323 \\
-0.323 & -0.323 & 0.598
\end{bmatrix}.
\end{align*}

\paragraph{Adversarial games.}
\begin{align*}
U_{\texttt{adv3\_01}} &=
\begin{bmatrix}
-0.152 & 1.186 & 0.482 \\
-0.250 & 0.000 & 0.392 \\
-0.537 & 0.530 & -1.650
\end{bmatrix},
&
U_{\texttt{adv3\_02}} &=
\begin{bmatrix}
-1.317 & -0.521 & 0.162 \\
0.160 & 0.078 & 0.820 \\
0.892 & 0.162 & -0.437
\end{bmatrix},
\\[1em]
U_{\texttt{adv3\_03}} &=
\begin{bmatrix}
-0.783 & -0.043 & 0.874 \\
-0.802 & 0.065 & -0.512 \\
1.316 & 0.291 & -0.405
\end{bmatrix},
&
U_{\texttt{adv3\_04}} &=
\begin{bmatrix}
-0.549 & 5.062 & -3.025 \\
-2.563 & 2.003 & -4.226 \\
1.512 & -1.086 & 2.873
\end{bmatrix}.
\end{align*}

\subsection{Sensitivity to the Number of Behavioral Prototypes $K$}
\label{app:k-sweep}

We vary the number of learned behavioral prototypes
$K_{\mathrm{model}}\in\{2,3,5\}$ while retaining three underlying
opponent regimes. All runs use the committed twelve-game $3\times3$
benchmark, ten seeds per game ($n=120$ matched game--seed cells), and
the headline evaluation settings: one focal agent with three persistent
opponents, $T=1000$, inclusive $H=100$, $\lambda=2$, $\alpha=0.9$,
and $M=2$. Only modeled prototype/belief cardinality varies; the
evaluation game suite, external switching protocol, and seed cells are
matched. The $K_{\mathrm{model}}=3$ run exactly reproduces the headline
PSR on all 120 matched cells.

Table~\ref{tab:k-sweep} reports absolute PSR and DPR. Relative to
$K_{\mathrm{model}}=3$, two prototypes increase mean PSR by $12.94$
(paired 95\% CI $[7.09,18.80]$; Holm-adjusted $p=5\times10^{-5}$),
although this penalty is not uniform across game families.
Five prototypes yield a $-1.32$ paired PSR difference
(95\% CI $[-4.94,2.30]$; Holm-adjusted $p=0.47$); this
interval includes zero. The corresponding DPR point estimates decrease
with modeled capacity, from $120.19$ to $106.67$ to $98.96$.
These results identify an aggregate under-specification cost in this
benchmark, but do not establish a PSR gain from over-specification or
explain how surplus prototypes are used. Game-clustered sensitivity
intervals also exclude zero for the two-prototype PSR contrast and
include zero for the five-prototype contrast.

\begin{table}[H]
\centering
\caption{Sensitivity to learned prototype count on the fixed twelve-game
$3\times3$ benchmark (three underlying behavioral regimes; ten seeds
per game). Entries are means with ordinary Student-$t$ 95\% CIs over
120 game--seed cells. Lower is better; paired PSR comparisons with
$K_{\mathrm{model}}=3$ are discussed in the text.}
\label{tab:k-sweep}
\begin{tabular}{lcc}
\toprule
\textbf{$K_{\mathrm{model}}$} & \textbf{PSR $\downarrow$} & \textbf{DPR $\downarrow$} \\
\midrule
2 & $175.91\;[158.03,193.79]$ & $120.19\;[116.96,123.43]$ \\
3 & $162.97\;[147.52,178.42]$ & $106.67\;[103.83,109.51]$ \\
5 & $161.65\;[146.05,177.24]$ & $98.96\;[96.18,101.74]$ \\
\bottomrule
\end{tabular}
\end{table}

\section{Mechanistic Analysis of Endogenous Adaptation}
\label{app:mechanistic}

This appendix examines regime-dependent adaptation under calibrated
behavioral shifts on the fixed twelve-game $3\times3$ benchmark. It
reports comparative PSR and changes in the oracle best response,
without isolating a causal mechanism.

\subsection{Behavioral Shift Difficulty}
\label{app:shift-difficulty}
We apply calibrated action-distribution shifts to the same twelve committed
$3\times3$ games. At $t=100$, each of three opponents changes from
$\pi_0=[0.8,0.1,0.1]$ to one of
$\pi_{\mathrm{small}}=[0.7,0.2,0.1]$,
$\pi_{\mathrm{medium}}=[0.5,0.3,0.2]$, or
$\pi_{\mathrm{large}}=[0.1,0.1,0.8]$.
We apply the same three cyclic action-label rotations to each game,
producing 36 game--rotation conditions at each shift magnitude, with ten
seeds per condition. A condition is decision-changing precisely when the
oracle row-player best-response index $\arg\max_i(U_{\mathrm{row}}\pi)_i$
differs before versus after the switch, using the same label rotation
and the lowest-index tie rule. The diagnostic separately measures behavioral
distance and whether a shift changes the decision-relevant oracle action.

\begin{table}[h]
\centering
\small
\caption{Calibrated $3\times3$ controlled shifts; TV and Jensen--Shannon (JS)
divergence between pre- and post-switch opponent action distributions.
Decision-changing conditions use the oracle row-player response.}
\label{tab:shift-calibration}
\begin{tabular}{lccc}
\toprule
Shift & TV & JS & Changed oracle action \\
\midrule
Small & $0.10$ & $0.010$ & $5/36$ ($13.9\%$) \\
Medium & $0.30$ & $0.052$ & $10/36$ ($27.8\%$) \\
Large & $0.70$ & $0.310$ & $26/36$ ($72.2\%$) \\
\bottomrule
\end{tabular}
\end{table}

\begin{table}[h]
\centering
\small
\caption{Mean PSR in the calibrated controlled-shift diagnostic
(lower is better); methods are not uniformly ranked across shifts.}
\label{tab:switch-difficulty}
\begin{tabular}{lcccc}
\toprule
Shift & ESL & Online EM & Fixed-$K$ & SA-FP \\
\midrule
Small & $\mathbf{14.17}$ & $14.67$ & $21.33$ & $26.09$ \\
Medium & $19.20$ & $\mathbf{18.37}$ & $19.77$ & $23.22$ \\
Large & $\mathbf{20.45}$ & $31.92$ & $50.85$ & $38.04$ \\
\bottomrule
\end{tabular}
\end{table}

ESL is not best at every shift magnitude: Online EM has the lowest mean PSR
under the medium shift, and ESL and Fixed-$K$ are similar there. The
largest ESL advantage in this diagnostic occurs under the large shift,
when $26/36$ conditions change the oracle best response. This suggests
that online prototype adaptation is especially useful in the evaluated
synthetic setting when shifts alter decision-relevant behavior; it does not
establish a universal relationship between shift size and adaptation.
% \subsection{Behavioral Shift Difficulty}

% \begin{table}[h]
% \centering
% \small
% \begin{tabular}{lccc}
% \toprule
% \textbf{Difficulty} & \textbf{ESL} & \textbf{Fixed-K} & \textbf{EM} \\
% \midrule
% Small  & 14.33 & \textbf{8.57}  & 11.37 \\
% Medium & 11.33 & \textbf{8.73}  & 10.30 \\
% Large  & \textbf{15.63} & 20.20 & 27.63 \\
% \bottomrule
% \end{tabular}
% \caption{
% Post-switch oracle regret (PSR) across controlled behavioral shifts (lower is better). ESL is not uniformly dominant: simpler methods perform competitively under small shifts, while ESL achieves the lowest regret under large behavioral changes requiring structural adaptation.
% }
% \label{tab:switch-difficulty}
% \end{table}

% ESL's advantage is regime-dependent. In low-difficulty regimes, where pre- and post-switch behaviors are similar, adaptation reduces to local probability calibration and simpler methods such as Fixed-K achieve lower regret. In contrast, ESL performs best under large behavioral shifts, where adaptation requires restructuring latent representations rather than incremental belief adjustment. This supports Proposition~\ref{prop:likelihood-mismatch}: likelihood-based methods can remain effective under locally stable interaction distributions while failing under substantial endogenous shifts.

\section{Diminishing-Step Diagnostic for Asymptotic Dynamics}
\label{app:rm-diagnostic}

The headline adaptation and ablation experiments use constant prototype
steps, small fixed batches ($M=2$), and no explicit prototype
regularization to study finite-horizon tracking. Theorem~\ref{thm:main-convergence}
instead concerns a diminishing-step, regularized recursion under a separate
fast-process averaging assumption.

We therefore run a \emph{finite-horizon diagnostic} on IPD, Stag Hunt,
and Matching Pennies (ten seeds each). Each run uses
\[
\gamma_m=0.5(m+1)^{-0.9},\qquad
M_m=\lceil50(m+1)^{0.2}\rceil,\qquad
\eta_{\mathrm{reg}}=10^{-3},
\]
for 80 completed slow updates (8,089 observed-record steps).
The step sizes satisfy the Robbins--Monro summability conditions.
Under the additional $O(M_m^{-1})$ tracking-bias bound discussed in
Appendix~\ref{app:convergence}, this schedule also makes
$\sum_m\gamma_m/M_m$ finite; however, the diagnostic does not
establish the frozen fast-process mixing condition required for that bound.

\begin{figure}[H]
\centering
\includegraphics[width=\linewidth]{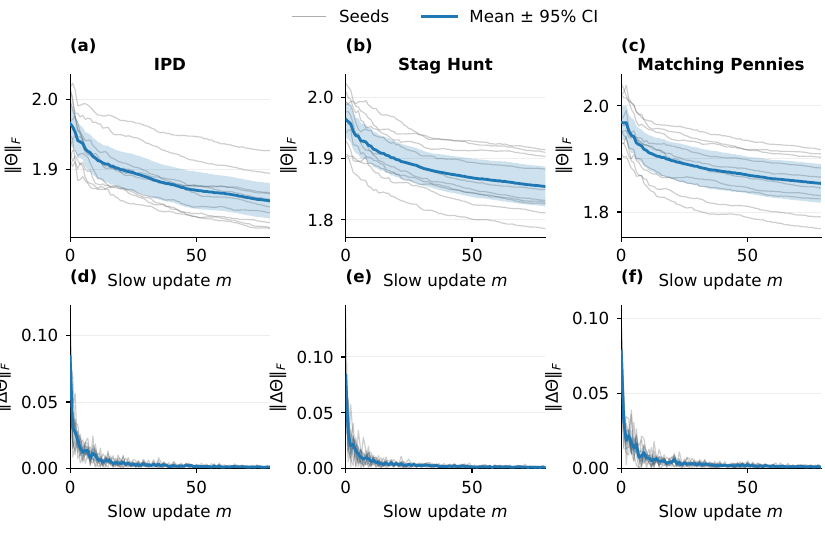}
\caption{
\textbf{Finite-horizon diminishing-step diagnostic.}
Across three two-action games (IPD, Stag Hunt, and Matching Pennies;
ten seeds per game), ESL uses diminishing steps
$\gamma_m=0.5(m+1)^{-0.9}$, growing batches
$M_m=\lceil50(m+1)^{0.2}\rceil$, and regularization
$\eta_{\mathrm{reg}}=10^{-3}$ for 80 slow updates.
Top panels: prototype parameter norms; bottom panels: update magnitudes.
Thin curves show individual runs; thick curves and shading show
per-game means and Student-$t$ 95\% confidence intervals across ten seeds.
These finite-horizon trajectories do not verify the averaging
assumption or establish convergence of the main constant-step algorithm.
}
\label{fig:rm-dynamics}
\end{figure}

Figure~\ref{fig:rm-dynamics} shows finite-horizon prototype norms of
approximately $1.77$--$2.04$ and declining update magnitudes across
the 30 game--seed runs. The latter partly reflects the prescribed
decrease in $\gamma_m$. Neither observation establishes asymptotic
boundedness, limiting internally chain-transitive behavior, or
Assumption~\ref{assum:fba}; this diagnostic is not a test or
verification of Theorem~\ref{thm:main-convergence}.

\section{Game-Theoretic Analysis of ESL}
\label{game-theoretic}

\paragraph{Game-theoretic interpretation of limit sets.}
The internally chain transitive (ICT) sets identified by
Theorem~\ref{thm:main-convergence} admit a natural game-theoretic
interpretation. Agents act through belief-conditioned logit best-response
policies and update their beliefs from realized interactions. The limiting
objects of ESL therefore do not necessarily coincide with Nash equilibria.
Instead, they are better understood as \emph{self-consistent behavioral
regimes} under bounded rationality~\cite{simon1955behavioral}, in which
beliefs, induced policies, prototype parameters, and the resulting data
distribution are mutually compatible under the endogenous learning dynamics.

The differential inclusion \eqref{eq:diff-inclusion} governs the slow-timescale
dynamics, and its ICT sets characterize invariant regimes of the system. Fixed
points correspond to stationary self-consistent configurations, while more
general ICT sets may capture recurrent, non-equilibrium behavior induced by
the belief--policy feedback loop. Thus, the theorem characterizes an
asymptotic ICT limit set, which need not be a single fixed point.

At a fixed point $\Theta^\ast$, there exists an invariant measure
$\mu^\ast \in \mathcal{I}(\Theta^\ast)$ such that
\[
g_{\mu^\ast}(\Theta^\ast)=0,
\]
meaning that the prototype parameters are stationary under a data distribution
that they themselves help induce. This admits an
\emph{Self-Confirming Equilibrium} (SCE)-like interpretation
\cite{fudenberg1993self}: beliefs need only be consistent with observations
generated along the realized path of play.
Nash equilibrium requires each player's strategy to be a best response
to opponents' actual strategies. By contrast, SCE permits best responses
to beliefs consistent with observed play, even when those beliefs are
incorrect about unobserved contingencies.

Because agents select actions through logit best responses, stationary ESL
configurations also admit a \emph{Quantal Response Equilibrium} (QRE)-like
interpretation~\cite{mckelvey1995quantal}. However, they are not generally QRE,
because responses may be based on learned, self-confirming beliefs rather than
the true strategy profile. Under additional conditions ensuring correct
beliefs, mutual consistency, and convergence to a stationary fixed point, the
connection to standard logit QRE becomes exact.

More general ICT sets need not admit a standard equilibrium interpretation.
Instead, they correspond to persistent behavioral regimes in which the system
moves through a sequence of mutually compatible configurations. This
highlights that endogenous interaction may generate non-convergent yet stable
learning dynamics.

\paragraph{Regret and strategic scope.}
ESL is not proved to achieve vanishing external or swap regret, and low PSR implies neither property. Conversely, low long-run regret can coexist with substantial regret immediately following a behavioral change, which is the local adaptation cost that PSR is designed to measure. Our experiments use committed switching policies rather than counterparts that adapt to ESL's learning state. We therefore make no robustness guarantee for such adaptive-counterpart settings.

\section{Representational and Computational Structure}
\label{app:complexity}

We distinguish storage for the shared behavioral representation from storage and computation associated with individual interactions.

\paragraph{Representation and belief storage.}
In the categorical model, the $K$ population-shared prototypes require $O(K|\mathcal A|)$ parameters, independent of $N$ for fixed $K$ and $|\mathcal A|$. Beliefs for $P$ tracked ordered agent pairs require an additional $O(PK)$ storage. Full directed-pair tracking has $P=N(N-1)$, whereas one focal agent tracking its opponents has $P=N-1$. Prototype sharing avoids a separate parameterized behavioral dictionary for every pair, but not pair-specific belief storage. The prototype count is fixed within each run and varied in Appendix~\ref{app:k-sweep}.

\paragraph{Per-batch computation.}
For a batch of $B_{\mathrm{rec}}$ observed records, direct dense evaluation of belief-weighted categorical prototype gradients takes $O(B_{\mathrm{rec}}K|\mathcal A|)$ operations. This is an implementation-level upper bound for the gradient calculation, not necessarily a tight bound for optimized aggregation. Bayesian belief updates, action selection, and interaction processing incur further work. Total computation thus depends on the number of encounters, not only on the shared prototype parameters. Neither total storage nor total computational cost is independent of population size.

\section{Runtime Diagnostic}
\label{app:runtime}

Prototype-gradient updates add per-batch work beyond belief inference;
Appendix~\ref{app:complexity} distinguishes this work from pairwise belief
storage and encounter processing.

We measure wall-clock time per run under the Figure~\ref{fig:dynamic-adaptation}
protocol (Table~\ref{tab:runtime-diagnostic}).\footnote{All measurements were
conducted on a MacBook Air (Apple M2, 16\,GB RAM, macOS Sequoia 15.5).}

\paragraph{Interpretation.}
ESL has lower measured wall-clock time than Online EM under this implementation
and protocol. This result is a finite-scale, CPU-specific diagnostic, not a
hardware-normalized comparison or evidence of superior asymptotic scaling.
Whether the ordering persists for larger populations, action spaces, or
other implementations remains untested.

\paragraph{Scaling behavior.}
\label{app:scaling}
The $O(B_{\mathrm{rec}}K|\mathcal A|)$ dense-gradient bound concerns one batch of observed
records, not the full algorithm or the population-wide interaction workload.
Shared prototypes reduce duplication of learned behavioral parameters;
pairwise beliefs and the number of processed encounters remain population
dependent. The opponent-count and recovery experiments
(Appendix~\ref{app:population-protocols}) do not establish
computational scalability.

\begin{table}[H]
\centering
\caption{
\textbf{Runtime diagnostic under the Figure 2 protocol.}
Mean wall-clock time per run (seconds) with 95\% confidence intervals, aggregated across tasks and seeds. Measurements reflect implementation-level timing on CPU and are not hardware-normalized benchmarks.
}
\label{tab:runtime-diagnostic}
\begin{tabular}{lcc}
\toprule
\textbf{Method} & \textbf{Wall-clock (s)} & \textbf{Relative vs.\ Online EM} \\
\midrule
ESL & $0.186 \; [0.184,\; 0.188]$ & $-62.6\%$ \\
Online EM & $0.497 \; [0.490,\; 0.504]$ & $0.0\%$ \\
\bottomrule
\end{tabular}
\end{table}

\paragraph{Compute resources and accounting.}
The reported experiments were run on CPU on a MacBook Air (Apple M2,
16\,GB RAM, macOS Sequoia 15.5); no GPU was used. Peak resident memory
was not instrumented, but the 16\,GB machine was sufficient. The public
camera-ready repository is approximately 5\,MB, including about
2.2\,MB of compact evidence and 112\,KB of fixed game files.
Optional, unpublished full population-recovery trajectories occupy about
1.4\,GB ($N=20$) and 4.7\,GB ($N=100$).
Table~\ref{tab:compute-accounting} distinguishes measured timing
records (M), proxies extrapolated from the measured $K=3$, $T=1000$
ESL cell mean of 1.20\,s (E), and unavailable timings (U).
The 0.186\,s ESL entry in Table~\ref{tab:runtime-diagnostic} is a
\emph{separate short Figure 2 diagnostic}, not the flagship cell time.

\begin{table}[H]
\centering
\footnotesize
\caption{\textbf{Indicative compute accounting.}
Cell counts are from the recorded evaluation configurations.
Hours for proxy rows are extrapolated per-cell wall-clock sums, not
measured CPU time; PSRV worker-hours assume all reported jobs were busy
for the entire recorded resumed session. U denotes unavailable timings.
Values are rounded and cover only evaluations reported in this paper.}
\label{tab:compute-accounting}
\begin{tabular}{@{}lrlr@{}}
\toprule
\textbf{Experiment family} & \textbf{Cells} & \textbf{Timing evidence} & \textbf{Hours} \\
\midrule
Flagship (incl. PPO) & 840 & E (PPO U) & $\sim0.3+\mathrm{U}$ \\
Matched ablations & 600 & E & $\sim0.20$ \\
Prototype coefficient & 600 & E & $\sim0.20$ \\
Batch size $M$ & 480 & E & $\sim0.16$ \\
Inverse temperature $\lambda$ & 600 & E & $\sim0.20$ \\
Prototype count $K$ & 360 & M (1.12\,s/cell mean) & $0.11$ \\
OMD/MWU, tune + evaluation & $\sim720$ & E & $\sim0.20$ \\
Opponent count $J=3,10,100$ & 360 & M (per-cell timers) & $0.59$ \\
Population recovery $N=20$ & 120 & M (0.22\,h session, 6 jobs) & $\sim1.3^{*}$ \\
Population recovery $N=100$ & 120 & M (9.43\,h session, 3 jobs) & $\sim28.3^{*}$ \\
Partial visibility & 3360 & E & $\sim1.1$ \\
Observation noise & 3360 & E & $\sim1.1$ \\
Controlled shifts & 5400 & E & $\sim1.8$ \\
Robbins--Monro diagnostic & 30 & M (17.6\,s job wall) & $0.005$ \\
Asymmetric-incentive suite & -- & U (run timing absent) & U \\
\bottomrule
\end{tabular}
\end{table}
\noindent\textit{Compute interpretation.} Under the stated proxy and job-occupancy
assumptions, the reconstructed reported-suite total is on the order of
$35$--$36$ worker-hours. This is \emph{not} a directly measured CPU-time
total, a lower bound, or a full-project accounting; PPO and asymmetric
runtime are unavailable, and actual PSRV worker utilization is unknown.
$^{*}$The PSRV figures multiply measured resumed-session wall time
by nominal job count.

\clearpage
\appendix

\section*{NeurIPS Paper Checklist}

\begin{enumerate}

\item {\bf Claims}
    \item[] Question: Do the main claims made in the abstract and introduction accurately reflect the paper's contributions and scope?
    \item[] Answer: \answerYes{} % Replace by \answerYes{}, \answerNo{}, or \answerNA{}.
    \item[] Justification: All the claims stated have been achieved and the contributions made by the paper have been clearly outlined in Section \ref{sec:intro}.

\item {\bf Limitations}
    \item[] Question: Does the paper discuss the limitations of the work performed by the authors?
    \item[] Answer: \answerYes{} % Replace by \answerYes{}, \answerNo{}, or \answerNA{}.
    \item[] Justification: The limitations are clearly outlined and discussed in Section \ref{sec:limitations}.

\item {\bf Theory assumptions and proofs}
    \item[] Question: For each theoretical result, does the paper provide the full set of assumptions and a complete (and correct) proof?
    \item[] Answer: \answerYes{} % Replace by \answerYes{}, \answerNo{}, or \answerNA{}.
    \item[] Justification: The theoretical assumptions for the results are outlined in Section \ref{sec:theory}. The section also accounts for a proof sketch which is formalized in the Appendix \ref{app:convergence}.

    \item {\bf Experimental result reproducibility}
    \item[] Question: Does the paper fully disclose all the information needed to reproduce the main experimental results of the paper to the extent that it affects the main claims and/or conclusions of the paper (regardless of whether the code and data are provided or not)?
    \item[] Answer: \answerYes{} % Replace by \answerYes{}, \answerNo{}, or \answerNA{}.
    \item[] Justification: The entire algorithm is given as Algorithm \ref{algo}. Aside from that, the appendix contains experimental details as Appendix \ref{sec:experimental_details} which includes hyperparameters in Appendix \ref{app:hyperparameters}. 

\item {\bf Open access to data and code}
    \item[] Question: Does the paper provide open access to the data and code, with sufficient instructions to faithfully reproduce the main experimental results, as described in supplemental material?
    \item[] Answer: \answerYes{} % Replace by \answerYes{}, \answerNo{}, or \answerNA{}.
    \item[] Justification: The public \href{https://github.com/jainendra2019/esl/releases/tag/neurips2026-flagship-cli}{ESL release} contains the implementation, frozen twelve-game payoff matrices, compact numerical evidence, dependencies, and reproduction commands. The README identifies the outputs reproduced from cell-level evidence and the secondary analyses available as precomputed summaries.

\item {\bf Experimental setting/details}
    \item[] Question: Does the paper specify all the training and test details (e.g., data splits, hyperparameters, how they were chosen, type of optimizer) necessary to understand the results?
    \item[] Answer: \answerYes{} % Replace by \answerYes{}, \answerNo{}, or \answerNA{}.
    \item[] Justification: Appendix~\ref{app:hyperparameters} gives the headline hyperparameters, observation and switching protocol, and baseline settings; Appendix~\ref{app:omd-protocol} documents the separate held-out tuning. The constant-step empirical regime and diminishing-step diagnostic are distinguished in Section~\ref{sec:theory} and Appendix~\ref{app:rm-diagnostic}.

\item {\bf Experiment statistical significance}
    \item[] Question: Does the paper report error bars suitably and correctly defined or other appropriate information about the statistical significance of the experiments?
    \item[] Answer: \answerYes{} % Replace by \answerYes{}, \answerNo{}, or \answerNA{}.
    \item[] Justification: The main twelve-game comparisons use ten seeds per game; relevant results report confidence intervals and paired tests over matched game--seed cells. Other diagnostics use their stated protocols, and deterministic calibration counts are not assigned sampling intervals. The aggregation and interval conventions are described in captions and Appendix~\ref{sec:experimental_details}.

\item {\bf Experiments compute resources}
    \item[] Question: For each experiment, does the paper provide sufficient information on the computer resources (type of compute workers, memory, time of execution) needed to reproduce the experiments?
    \item[] Answer: \answerYes{} % Replace by \answerYes{}, \answerNo{}, or \answerNA{}.
    \item[] Justification: Appendix~\ref{app:runtime} documents the CPU-only Apple M2 environment, memory and storage needs, and per-family evaluation counts with measured, estimated, or unavailable timings (Table~\ref{tab:compute-accounting}). The reconstructed reported-suite effort is on the order of tens of worker-hours under stated assumptions, not measured total CPU time. The appendix distinguishes the short runtime diagnostic from flagship evaluations. The total is estimated for reported experiments only; complete project-level compute was not measured.
    
\item {\bf Code of ethics}
    \item[] Question: Does the research conducted in the paper conform, in every respect, with the NeurIPS Code of Ethics \url{https://neurips.cc/public/EthicsGuidelines}?
    \item[] Answer: \answerYes{} % Replace by \answerYes{}, \answerNo{}, or \answerNA{}.
    \item[] Justification: There were no harms in the research process along with no data related concerns as we use theoretical experiments. There are no dual-use risks posed by the same. The authors have reviewed the NeurIPS Code of Ethics and confirm compliance in all respects.

\item {\bf Broader impacts}
    \item[] Question: Does the paper discuss both potential positive societal impacts and negative societal impacts of the work performed?
    \item[] Answer: \answerNA{} % Replace by \answerYes{}, \answerNo{}, or \answerNA{}.
    \item[] Justification: This paper presents foundational algorithmic research on multi-agent learning dynamics in synthetic repeated game    settings. It does not introduce a deployable system, a new dataset, or a model trained on real-world data, and it has no direct path to specific harmful applications. The methods developed --- Bayesian     belief updating over shared latent prototypes, controlled-Markov stochastic approximation --- are standard tools in the multi-agent learning literature.
    
\item {\bf Safeguards}
    \item[] Question: Does the paper describe safeguards that have been put in place for responsible release of data or models that have a high risk for misuse (e.g., pre-trained language models, image generators, or scraped datasets)?
    \item[] Answer: \answerNA{} % Replace by \answerYes{}, \answerNo{}, or \answerNA{}.
    \item[] Justification: The paper releases no pre-trained models, image generators, or scraped datasets. All experiments are conducted on synthetic repeated matrix games with no connection to personal data. The released code consists of simulation environments and learning algorithm implementations that pose no meaningful misuse risk.

\item {\bf Licenses for existing assets}
    \item[] Question: Are the creators or original owners of assets (e.g., code, data, models), used in the paper, properly credited and are the license and terms of use explicitly mentioned and properly respected?
    \item[] Answer: \answerYes{} % Replace by \answerYes{}, \answerNo{}, or \answerNA{}.
    \item[] Justification: The synthetic game benchmark and ESL code are released under the repository MIT license. The optional vendored PPO implementation retains its MIT attribution; relevant prior methods and theoretical tools are cited. No proprietary datasets are used.

\item {\bf New assets}
    \item[] Question: Are new assets introduced in the paper well documented and is the documentation provided alongside the assets?
    \item[] Answer: \answerYes{} % Replace by \answerYes{}, \answerNo{}, or \answerNA{}.
    \item[] Justification: The \href{https://github.com/jainendra2019/esl/releases/tag/neurips2026-flagship-cli}{ESL release} provides the algorithm, evaluated baselines, twelve fixed $3\times3$ games across three game families, configurations, metrics, and numerical evidence. Its README documents installation and reproduction; the figure script produces supporting evidence-derived plots rather than exact manuscript figure PDFs.

\item {\bf Crowdsourcing and research with human subjects}
    \item[] Question: For crowdsourcing experiments and research with human subjects, does the paper include the full text of instructions given to participants and screenshots, if applicable, as well as details about compensation (if any)? 
    \item[] Answer: \answerNA{} % Replace by \answerYes{}, \answerNo{}, or \answerNA{}.
    \item[] Justification: This paper involves no crowdsourcing and no research with human subjects. All experiments are conducted entirely in simulation using synthetic multi-agent environments.

\item {\bf Institutional review board (IRB) approvals or equivalent for research with human subjects}
    \item[] Question: Does the paper describe potential risks incurred by study participants, whether such risks were disclosed to the subjects, and whether Institutional Review Board (IRB) approvals (or an equivalent approval/review based on the requirements of your country or institution) were obtained?
    \item[] Answer: \answerNA{} % Replace by \answerYes{}, \answerNo{}, or \answerNA{}.
    \item[] Justification: No human subjects are involved at any stage of this research and all experiments are fully synthetic.

\item {\bf Declaration of LLM usage}
    \item[] Question: Does the paper describe the usage of LLMs if it is an important, original, or non-standard component of the core methods in this research? Note that if the LLM is used only for writing, editing, or formatting purposes and does \emph{not} impact the core methodology, scientific rigor, or originality of the research, declaration is not required.
    %this research? 
    \item[] Answer: \answerNA{} % Replace by \answerYes{}, \answerNo{}, or \answerNA{}.
    \item[] Justification: LLMs are not used in any component of the core methodology of the paper. Any use of the same was limited to writing assistance and does not affect the scientific content of the paper.

\end{enumerate}

%%%%%%%%%%%%%%%%%%%%%%%%%%%%%%%%%%%%%%%%%%%%%%%%%%%%%%%%%%%%

% \newpage
% \input{checklist.tex}

\end{document}